\documentclass[manuscript,nonacm]{acmart}

\usepackage{booktabs}
\usepackage{multirow}
\usepackage{array}
\usepackage{tabularx}
\usepackage{longtable}
\usepackage{xcolor}
\usepackage{enumitem}

\setcopyright{none}

\renewcommand{\department}[2][0]{\unskip~#2,\ignorespaces}

\begin{document}

\title{On Edge in the Dental Chair: Designing VR Support for Moments of Dental Anxiety}

% 1. Zhu Guo
\author{Zhu Guo}
\authornote{These authors contributed equally to this work.}
\orcid{0009-0008-0093-9188}
\affiliation{%
  \department{School of Artificial Intelligence}
  \institution{The Chinese University of Hong Kong, Shenzhen}
  \city{Shenzhen}
  \country{China}
}
\email{guozhu@link.cuhk.edu.cn}

% 2. Junjie Zhao
\author{Junjie Zhao}
\authornotemark[1]
\affiliation{%
  \department[0]{Division of Applied Oral Sciences and Community Dental Care}
  \department[1]{Faculty of Dentistry}
  \institution{The University of Hong Kong}
  \city{Hong Kong}
  \country{China}
}
\affiliation{%
  \institution{Beijing Institute of Collaborative Innovation}
  \city{Beijing}
  \country{China}
}
\email{junjiezhao@connect.hku.hk}

% 3. Haofan He
\author{Haofan He}
\affiliation{%
  \department{School of Design}
  \institution{Hunan University}
  \city{Changsha}
  \country{China}
}
\email{hehaofan@hnu.edu.cn}

% 4. Jiaming Zhang
\author{Jiaming Zhang}
\affiliation{%
  \department[0]{Shenzhen Clinical College of Stomatology}
  \department[1]{School of Stomatology}
  \institution{Southern Medical University}
  \city{Shenzhen}
  \country{China}
}
\affiliation{%
  \department{Department of Periodontology}
  \institution{Shenzhen Stomatology Hospital (Pingshan) of Southern Medical University}
  \city{Shenzhen}
  \country{China}
}
\email{710455941@qq.com}

% 5. Mingshi Deng
\author{Mingshi Deng}
\affiliation{%
  \department{School of Science and Engineering}
  \institution{The Chinese University of Hong Kong, Shenzhen}
  \city{Shenzhen}
  \country{China}
}
\email{mingshideng@link.cuhk.edu.cn}

% 6. Mingjun Zhou
\author{Mingjun Zhou}
\affiliation{%
  \department{School of Data Science}
  \institution{The Chinese University of Hong Kong, Shenzhen}
  \city{Shenzhen}
  \country{China}
}
\email{124090989@link.cuhk.edu.cn}

% 7. Dongyijie Primo Pan
\author{Dongyijie Primo Pan}
\affiliation{%
  \department{Computational Media and Arts}
  \institution{The Hong Kong University of Science and Technology (Guangzhou)}
  \city{Guangzhou}
  \country{China}
}
\email{dpan750@connect.hkust-gz.edu.cn}

% 8. Zikun Jin
\author{Zikun Jin}
\affiliation{%
  \department[0]{Division of Applied Oral Sciences and Community Dental Care}
  \department[1]{Faculty of Dentistry}
  \institution{The University of Hong Kong}
  \city{Hong Kong}
  \country{China}
}
\email{u3015427@connect.hku.hk}

% 9. Jianquan Li
% Author confirmation pending: institutional name and email (carried over from source).
\author{Jianquan Li}
\affiliation{%
  \institution{Shenzhen FreedomAI Technology Inc.}
  \city{Shenzhen}
  \country{China}
}
\email{lijianquan@freedomai.cn}

% 10. Liangyi Chen
% The supplied Peking University unit list is preserved without inferring internal hierarchy.
\author{Liangyi Chen}
\affiliation{%
  \department{New Cornerstone Science Laboratory, National Biomedical Imaging Center, State Key Laboratory of Membrane Biology, Institute of Molecular Medicine, Peking-Tsinghua Center for Life Sciences, College of Future Technology, IDG/McGovern Institute for Brain Research}
  \institution{Peking University}
  \city{Beijing}
  \country{China}
}
\affiliation{%
  \institution{Beijing Institute of Collaborative Innovation}
  \city{Beijing}
  \country{China}
}
\email{lychen@pku.edu.cn}

% 11. Zuolin Jin
% Author confirmation pending: applicable affiliation (carried over from source).
\author{Zuolin Jin}
\affiliation{%
  \institution{Shenzhen Stomatology Hospital (Pingshan) of Southern Medical University}
  \city{Shenzhen}
  \country{China}
}
\email{zuolinj@163.com}

% 12. Benyou Wang
\author{Benyou Wang}
\authornote{Corresponding authors: Benyou Wang, Jie Li, and Siying Hu.}
\affiliation{%
  \department{School of Data Science}
  \institution{The Chinese University of Hong Kong, Shenzhen}
  \city{Shenzhen}
  \country{China}
}
\affiliation{%
  \institution{Shenzhen Loop Area Institute}
  \city{Shenzhen}
  \country{China}
}
\email{wangbenyou@cuhk.edu.cn}

% 13. Jie Li
\author{Jie Li}
\authornotemark[2]
\affiliation{%
  \department{MIT Media Lab}
  \institution{Massachusetts Institute of Technology}
  \city{Cambridge}
  \state{Massachusetts}
  \country{USA}
}
\email{jieli8@mit.edu}

% 14. Siying Hu
\author{Siying Hu}
\authornotemark[2]
\orcid{0000-0002-3824-2801}
\affiliation{%
  \department{School of Computing}
  \institution{The Australian National University}
  \city{Canberra}
  \country{Australia}
}
\email{sying.ch1026@gmail.com}

% 15. Shan Jiang
\author{Shan Jiang}
\affiliation{%
  \department{Department of Periodontology}
  \institution{Shenzhen Stomatology Hospital (Pingshan) of Southern Medical University}
  \city{Shenzhen}
  \country{China}
}
\email{jshan@smu.edu.cn}

% 16. Junwen Wang
\author{Junwen Wang}
\affiliation{%
  \department[0]{Division of Applied Oral Sciences and Community Dental Care}
  \department[1]{Faculty of Dentistry}
  \institution{The University of Hong Kong}
  \city{Hong Kong}
  \country{China}
}
\affiliation{%
  \institution{Beijing Institute of Collaborative Innovation}
  \city{Beijing}
  \country{China}
}
\email{junwen@hku.hk}

\renewcommand{\shortauthors}{Guo et al.}

\begin{abstract}
Dental anxiety can change as a procedure unfolds, yet dental virtual
reality (VR) commonly provides continuous distraction or relaxation.
We investigate how support can be coordinated with specific simulated
dental events. Stakeholder interviews ($N=36$), participatory design
with three returning dentists, and patient walkthroughs of a
no-intervention prototype ($N=12$) informed five Anxiety Events and
an intervention-module framework. Drawing on cognitive vulnerability
and emotion regulation, we implemented a standardized event-contingent
VR system with predefined event--module assignments and shared agency
and safety controls. A randomized study ($N=24$) compared the
intervention package with no-intervention VR. The adjusted
intervention-minus-control difference averaged $-12.83$ VAS-A points
across events (95\% CI $[-24.42,-1.70]$). Physiological, behavioural,
and qualitative measures contextualized participants' experiences.
The findings inform timely, comprehensible support and reassuring
social presence in simulated dental VR; they concern the complete
package rather than individual modules or clinical effectiveness.
\end{abstract}




\maketitle

\section{Introduction}

Dental anxiety can make dental care distressing and contribute to avoidance, with an estimated 15.3\% of adults experiencing dental fear or anxiety worldwide~\cite{silveira2021prevalence,hoffmann2022management}. Invasive procedures are particularly challenging because anticipated pain, instruments, unfamiliar sensations, and previous negative experiences can make treatment threatening~\cite{astramskaite2016extraction,oosterink2008stimuli}. During third-molar extraction under local anaesthesia, patients remain conscious as the procedure progresses through preparation, instrument approach, drilling, and continued treatment~\cite{yamashita2020vr}. These moments can create different support needs, even within one appointment. For HCI, the challenge is to coordinate support with an unfolding procedure while preserving patients' understanding and control.

\begin{figure*}[t]
  \centering
  \includegraphics[width=\textwidth]{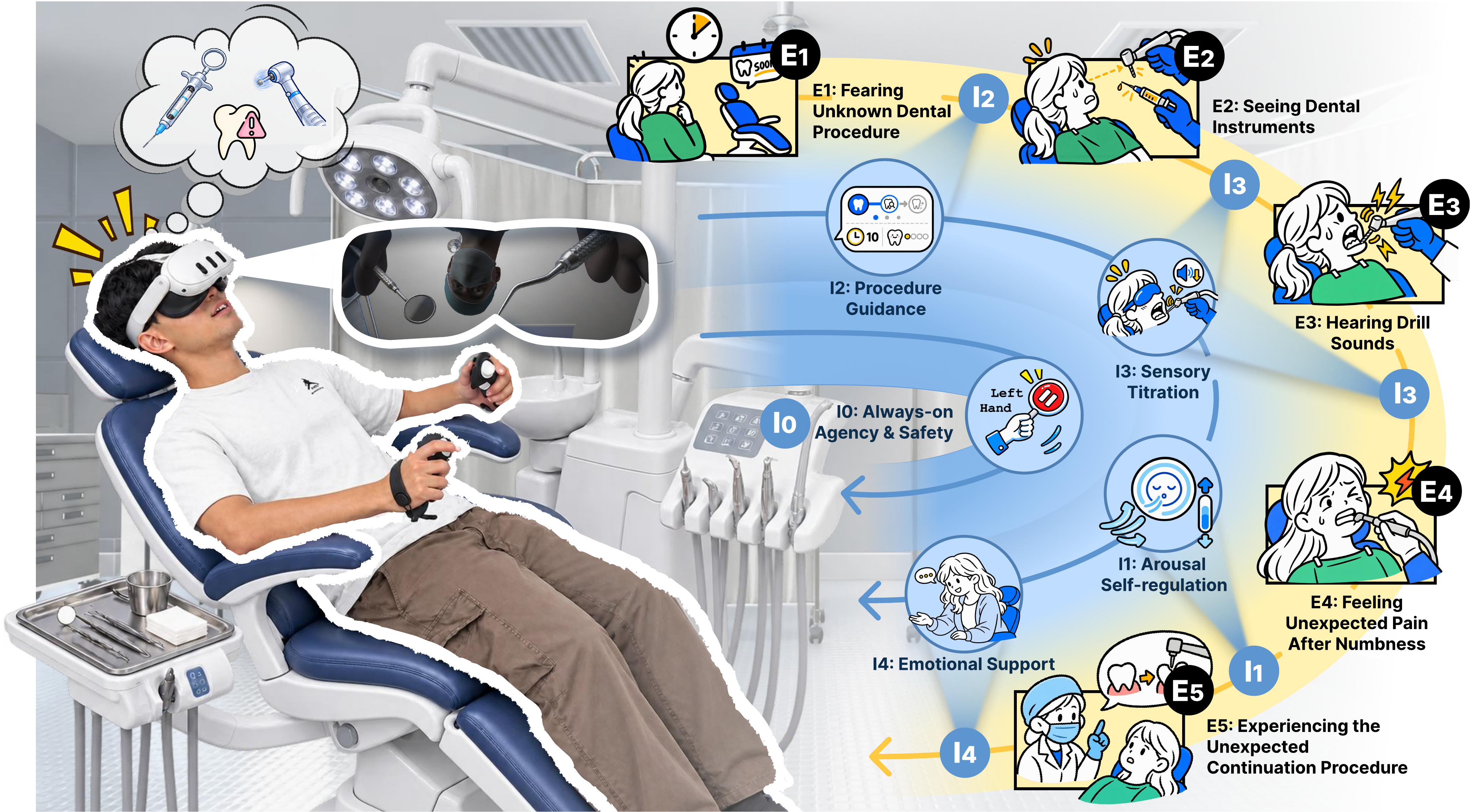}
  \caption{Overview of the standardized event-contingent VR system.
The left side illustrates the patient context and the view presented
in VR. On the right, three curved arrows schematically represent
the temporal progression of anxiety events and intervention delivery.
Black markers denote anxiety events (E1--E5), and blue markers denote
interventions (I1--I4), following the fixed sequence:
E1 $\rightarrow$ I2 $\rightarrow$ E2 $\rightarrow$ I3
$\rightarrow$ E3 $\rightarrow$ I3 $\rightarrow$ E4
$\rightarrow$ I1 $\rightarrow$ E5 $\rightarrow$ I4.
Procedure guidance (I2) follows E1; visual and auditory sensory
titration (I3) follow E2 and E3, respectively; arousal self-regulation
(I1) follows E4; and emotional support (I4) follows E5.
The agency and safety layer (I0) remains available throughout
the sequence, independently of scheduled intervention delivery,
allowing participants to pause, resume, or exit.
The arrows indicate order without encoding event or intervention
duration.}
  \Description{A three-part overview shows a participant in the simulated dental setting, the system's intervention components, and a five-event trajectory with predefined intervention delivery.}
  \label{fig:system-teaser}
\end{figure*}

Dental-anxiety management includes cognitive behavioural interventions, relaxation, procedural information, patient-control mechanisms, distraction, and sedation~\cite{hoffmann2022management,steenen2024interventions}. Virtual reality (VR) extends this repertoire through immersive distraction, calming environments, and controlled exposure~\cite{lopezvalverde2020vr}. However, evidence for adult state-anxiety benefits remains heterogeneous~\cite{steenen2024interventions}. Evaluations of continuous VR content and session-level outcomes leave a more specific design question unresolved: how should support relate to particular threats as a procedure unfolds? An explanation may address uncertainty before treatment, whereas a later sensory cue or violated expectation may require another form of support. Aggregate outcomes provide limited guidance for examining these differences or coordinating intervention timing with the procedural context.

We investigate this question through a simulated dental procedure with predefined Anxiety Events. Our system delivers a standardized intervention at each event and preserves an always-available agency and safety layer. We call this approach \emph{event-contingent}: delivery follows a fixed event script rather than an inferred user state. The Cognitive Vulnerability Model (CVM) informs our interpretation of threat appraisals~\cite{armfield2006cvm,armfield2008dental}; emotion-regulation theory informs the intended roles of support~\cite{gross1998emotion,zaki2013interpersonal}. These perspectives connect stakeholder accounts to design decisions without assuming that theoretical correspondence establishes intervention effectiveness. Figure~\ref{fig:system-teaser} provides an overview of the resulting system.

Three research questions organize the study, from understanding support needs to implementing and evaluating the intervention package:

\begin{quote}
\textbf{RQ1:} What threats characterize patients' dental anxiety, and what support strategies do stakeholders consider appropriate for addressing them?

\textbf{RQ2:} How can these threats and support strategies inform an event-contingent VR system that preserves patient agency and procedural comprehensibility?

\textbf{RQ3:} How does adding the standardized intervention package affect self-reported anxiety during simulated dental events, and how do physiological, behavioural, and qualitative measures contextualize participants' experiences?
\end{quote}

To address RQ1, we interviewed 36 stakeholders: 12 patients, 12 dental students, and 12 healthcare professionals. Three returning dentists then translated the findings into a third-molar-extraction scenario, preliminary events, and an intervention-module framework. A walkthrough with 12 patients examined a common no-intervention prototype and informed refinements to event presentation and transitions. These phases supplied complementary evidence about support needs, expert design judgments, and patients' interpretations of the simulated sequence.

For RQ2, we implemented five Anxiety Events and four event-specific modules: arousal self-regulation, procedure guidance, sensory titration, and emotional support. Each event invokes one predefined module, while the shared agency and safety layer remains available throughout. The design account traces threats and support needs through formative decisions to the implemented event--module assignments. It also distinguishes expert recommendations from later research-team choices, including the final fixed mapping.

For RQ3, a two-arm parallel randomized user study compared the intervention package with no-intervention VR ($N=24$, 12 per condition). Both conditions shared the event sequence, equipment, instructions, event-specific measurements, and baseline safety and withdrawal functions. Runs lasted approximately 15 minutes with interventions and 10 minutes without them; the control included no additional waiting to equalize duration. Event-specific anxiety ratings were the primary outcome. Physiological arousal, head-directed visual orientation, interaction records, and interviews provided complementary evidence. The emotional-support character appeared only during E5 in the intervention condition and offered spoken companionship; no corresponding character appeared in the control. The comparison estimates the immediate effect of adding the complete package within the simulated sequence. It cannot separate individual module effects or establish the advantage of event-contingent timing over another delivery schedule.

Our contributions are:

\begin{itemize}
    \item \textbf{An empirically grounded design account of dental-anxiety support.} Three formative phases connect recurring threats and stakeholder support needs to a bounded scenario, event representations, and four design goals.
    \item \textbf{A standardized event-contingent VR system.} The implementation operationalizes these findings through predefined event--module assignments and an independently available agency and safety layer.
    \item \textbf{An event-level evaluation and design implications.} A randomized study combines self-reported anxiety with physiological, behavioural, and qualitative evidence to examine the package and derive implications for comprehensible, minimally disruptive support.
\end{itemize}

\section{Related Work}

\subsection{Dental Anxiety and Emotion Regulation}\label{sec:rw-dental-anxiety}

Dental anxiety encompasses emotional, cognitive, physiological, and behavioural responses to dental care. Avoidance can contribute to deteriorating oral health and more invasive treatment, reinforcing anxiety~\cite{armfield2013cycle}. Patients describe anticipated pain, uncertainty, perceived lack of control, and previous negative encounters as sources of distress~\cite{wang2017patientviews,hoffmann2022management}. During conscious treatment, these concerns can change as instruments appear, sensations begin, or a procedure continues longer than expected. Understanding the appraisal associated with each moment can therefore inform what support should address.

The Cognitive Vulnerability Model (CVM) characterizes fear through appraisals of dangerousness, uncontrollability, unpredictability, and related vulnerability perceptions~\cite{armfield2006cvm}. Dental research associates cognitive vulnerability with fear while showing that its components can contribute differently across people and situations~\cite{armfield2008dental}. For interaction design, this framework directs attention to how information, control, sensory cues, and expectations shape a threatening experience. We use it to connect stakeholder accounts with the aspects of a dental event that support might address.

Management strategies include relaxation, information, patient-control mechanisms, distraction, supportive communication, and cognitive or behavioural approaches~\cite{hoffmann2022management,steenen2024interventions}. Patients also value reassurance, clear explanations, and opportunities to signal discomfort~\cite{wang2017patientviews}. Gross's Process Model of Emotion Regulation distinguishes strategies by their position within emotion generation~\cite{gross1998emotion}. Situation modification changes the circumstances; attentional deployment redirects attention; cognitive change alters interpretation; and response modulation acts on an emerging response. Procedural guidance may support reinterpretation, while breathing guidance may help regulate arousal. Interpersonal emotion regulation further explains how another social actor can provide emotional support~\cite{zaki2013interpersonal}. These perspectives help distinguish an event's threat appraisal from the regulatory action intended to address it.

\subsection{Cognitive Behavioural Approaches and Exposure Learning}\label{sec:rw-cbt-exposure}

Cognitive behavioural therapy (CBT) links a person's interpretations, emotions, and actions within an individualized account of what maintains their difficulties. Its practices can include examining distressing thoughts, testing predictions through behaviour, and rehearsing coping responses~\cite{beck2021basics}. This perspective complements CVM: CVM identifies vulnerability-related appraisals, whereas CBT asks how interpretations and responses might be examined and changed. A dental explanation can supply information relevant to a threat belief, but a structured cognitive exercise additionally requires the person to identify their interpretation and consider what supports or challenges it. The presence of information or a relaxation cue alone does not specify that therapeutic sequence.

HCI research makes this distinction concrete. Kitson et al. identify challenges in recognizing thoughts, generating alternative appraisals, and applying reappraisal during emotionally demanding situations~\cite{kitson2024reappraisal}. Burger et al.'s conversational thought-recording study examines a structured activity through participants' records, feedback, and self-reflection~\cite{burger2022thoughtrecord}. These works suggest that support should be described through what a person can do with it. For dental VR, relevant questions include whether guidance helps someone articulate what they expect, whether they can follow a coping activity while the procedure continues, and what opportunity they have to reflect afterwards. These are different interaction requirements from delivering a reassuring message.

Exposure-based approaches add a learning objective to encountering feared situations. In an inhibitory-learning account, exposure provides opportunities to develop competing associations that can be retrieved when fear returns; learning is not indexed simply by becoming calmer within a session~\cite{craske2014inhibitory}. A central design question is which feared prediction the experience permits the person to test. This therapeutic use of expectancy violation differs from introducing an unexpected aversive event: a discrepancy can strengthen a threat belief as well as challenge it. The distinction matters for a dental simulation in which an unexpected sensation or additional operation may itself be the source of anxiety.

Support during exposure also has to be understood by its function. A person may use an aid to remain engaged with a difficult situation, or may conclude that the feared outcome was prevented only by that aid. Research therefore examines the conditions under which safety behaviours help or interfere with exposure learning. In a randomized spider-fear study, Blakey et al. found no significant outcome or acceptability differences between eliminating safety behaviours and introducing them judiciously before fading them~\cite{blakey2019safety}. That result does not settle their role in dental VR, but cautions against classifying all comforting features as either beneficial or countertherapeutic. Their use, interpretation, and effects on later engagement need to be examined.

\subsection{Virtual Reality for Dental Anxiety}\label{sec:rw-dental-vr}

Dental VR includes immersive distraction, calming audiovisual content, and controlled exposure to feared situations~\cite{lopezvalverde2020vr,nezhad2024vr,yang2025xr}. Distraction redirects attention or reduces access to procedure-related cues; exposure retains or reproduces feared cues within a therapeutic exercise. These approaches address different goals. Reviews report variation across content, procedure, population, and comparison condition~\cite{lopezvalverde2020vr,nezhad2024vr,yang2025xr}. Benefits from paediatric distraction studies cannot be assumed to transfer to adult invasive care.

Two lines of adult dental research illustrate the distinction. Yamashita et al. evaluated VR during impacted mandibular third-molar extraction under local anaesthesia~\cite{yamashita2020vr}. Gujjar et al. instead randomized adults with dental phobia to VR exposure therapy or an informational pamphlet, assessing anxiety over follow-up alongside behavioural avoidance and later treatment acceptance~\cite{gujjar2019dental}. The former situates support within ongoing care; the latter evaluates a therapeutic intervention intended to change the person's relationship to future dental care. Thus, immediate procedural anxiety, avoidance, and longer-term dental fear are related but distinct targets for design and evaluation.

VR exposure research beyond dentistry also shows how therapeutic work is organized through interaction. Brinkman et al. redesigned a fear-of-flying therapist interface around treatment scenarios rather than the manual operation of simulation controls~\cite{brinkman2010therapist}. Freeman et al. evaluated an automated cognitive intervention for fear of heights in which a virtual coach guided therapeutic tasks across repeated sessions~\cite{freeman2018automated}. These examples foreground the selection and guidance of activities, as well as who controls their progression. Their clinical results do not transfer directly to dentistry, but their interaction arrangements help distinguish a simulated encounter, a supported encounter, and a structured therapeutic exercise.

Our study examines support within a fixed dental-event sequence. It preserves the procedural scene while adding explanations, selective sensory attenuation, breathing guidance, and companionship at predefined moments. This makes it possible to examine how support is experienced as the procedure unfolds. The system was evaluated as an immediate-support package, without an individualized exposure hierarchy, explicit prediction-testing exercises, or follow-up assessment of learning. Its relevance to CBT and exposure therefore lies in the design questions these approaches reveal, rather than evidence that the package constitutes or achieves their therapeutic effects.

\subsection{User-State Assessment and Timely Intervention in HCI}\label{sec:rw-hci-support}

HCI offers complementary approaches to measuring user experience and delivering support at relevant moments. Physiological computing treats bodily signals as a channel for understanding or adapting interaction~\cite{fairclough2009physiological}. Heart activity and electrodermal activity (EDA) can be recorded continuously, making them useful for event-related analysis. Yet autonomic activation varies across emotions, individuals, and contexts, so these signals are not uniquely specific to anxiety~\cite{kreibig2010autonomic}. EDA reflects sympathetic influence on sweat-gland activity and includes tonic and phasic components; acquisition choices and movement also affect its interpretation~\cite{posada2020eda}. We consequently describe heart rate (HR) and EDA as measures of physiological arousal.

Head orientation offers a behavioural account of how users direct the headset within a virtual scene. A forward ray can indicate whether a predefined scene region lies near the centre of the view. However, visual exploration combines head movement and ocular gaze, and these are not equivalent~\cite{sitzmann2018saliency}. Head orientation cannot establish ocular fixation or detect attention shifts without corresponding movement. In our study, \emph{head-directed visual orientation} therefore complements self-report without serving as a direct measure of gaze or anxiety.

These distinctions support multimodal assessment in which each measure retains a defined role. Event-specific self-report records experienced anxiety; physiological signals characterize arousal; orientation and interaction records describe observable behaviour; and interviews contextualize participants' interpretations. Aligning these sources with events improves temporal resolution without making them interchangeable. Our physiological and orientation data are analysed offline and do not trigger intervention delivery.

These measurement and timing questions connect to theory-informed intervention design. Slovak et al. distinguish intended change, intervention mechanisms, and their technological implementations; Slovak and Munson further relate design contributions at different scopes to the evidence needed to assess them~\cite{slovak2023emotion,slovak2024framework}. In this view, a timed prompt is an implementation choice whose value depends on the activity it enables. In dental VR, this means specifying whether an event-linked explanation is intended to inform, whether a breathing cue enables practice, or whether a companion offers social support, and evaluating those functions alongside the overall experience.

Just-in-time adaptive interventions specify decision points, intervention options, tailoring variables, and decision rules~\cite{nahumshani2018jitai}. Related VR work has combined real-time user-state inference with support during acute cognitive overload~\cite{zhang2026seconds}. Such approaches depend on reliable sensing, inference, calibration, and delivery logic. A mistimed or opaque intervention may disrupt the experience or undermine perceived control, particularly in a sensitive medical simulation. State-based delivery therefore requires evidence that the inferred state is reliable enough for the decisions it informs.

Event-contingent delivery provides a design alternative when a procedure has known stages and real-time anxiety inference has not been established. Synchronizing support with predefined events standardizes timing across participants and makes delivery rules available for inspection. It still requires decisions about which threat matters, when support should appear, and how to preserve necessary information and safety controls. We examine these interaction-design decisions through a fixed intervention script. The resulting system is event-contingent, rather than adaptive or a JITAI, and its evaluation concerns the complete scripted package.

\section{Formative Study}
\label{sec:formative}

\begin{figure*}[t]
  \centering
  \includegraphics[width=\textwidth]{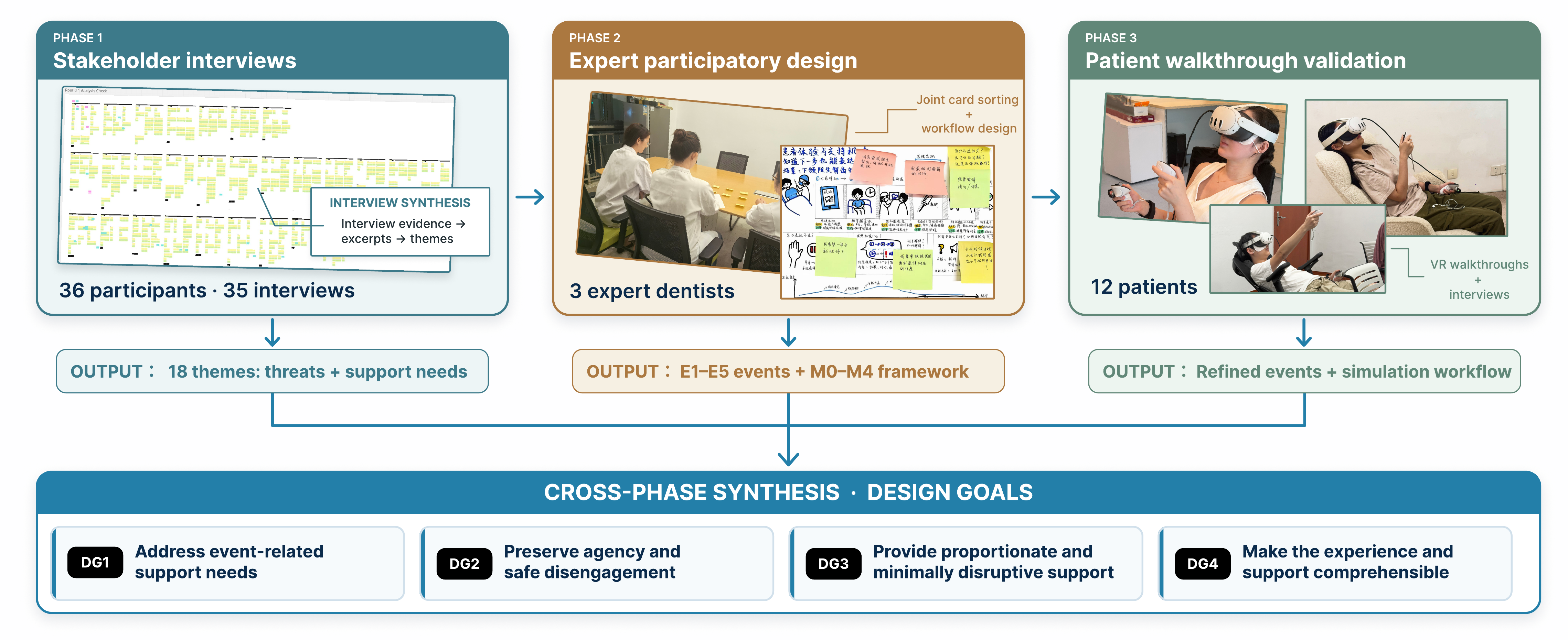}
  \caption{Overview of the three-phase formative study.
  Phase~1 involved 36 stakeholders through 35 synchronous
  interviews and one written response.
  Phase~2 involved three returning dentists in participatory
  design, producing preliminary E1--E5 events and an M0--M4
  module framework.
  Phase~3 involved 12 patients in a common no-intervention
  VR walkthrough and interviews, informing subsequent event
  and workflow refinements.
  The cross-phase synthesis informed DG1--DG4.
  Phase-specific samples overlap; Phase~3 did not evaluate
  M1--M4 or retest the subsequent revisions.}
  \Description{Three columns show an interview-analysis board,
  expert card sorting and an annotated design sheet, and
  photographs of patient VR walkthroughs. Output boxes beneath
  the columns summarize themes, preliminary events and modules,
  and event and workflow refinements. Arrows connect these
  outputs to four design goals concerning event-related support,
  agency, minimally disruptive support, and comprehensibility.}
  \label{fig:formative-process}
\end{figure*}

We conducted a three-phase formative study to answer RQ1 and to establish the design basis for the subsequent VR system. Phase~1 used stakeholder interviews to identify recurring \emph{Anxiety Threats}, candidate intervention approaches, and cross-cutting system considerations. Phase~2 used expert participatory design to translate these findings into a bounded third-molar-extraction scenario, five preliminary \emph{Anxiety Events}, an M0--M4 intervention-module framework, and a preliminary workflow. Phase~3 exposed patients to a common no-intervention VR prototype to examine whether the proposed event sequence was plausible, comprehensible, tolerable under supervision, and in need of refinement. Table~\ref{tab:formative-overview} summarizes the role of each phase; detailed recruitment, interview materials, coding procedures, full theme inventories, expert refinements, prototype version history, and evidence boundaries are reported in Appendix~\ref{app:formative-details}. Figure~\ref{fig:formative-process} illustrates the formative
process and its synthesis into design goals.

Participants in Phases~1 and~2 were recruited through a university-affiliated dental hospital in southern China; Phase~3 participants were recruited from the local community and a university campus. The formative-study protocol was approved by the Ethics Committee of The Chinese University of Hong Kong, Shenzhen (approval no.~CUHKSZ-D-20260095) before data collection. Participants were informed of the study purpose and intended research use of their data and provided oral informed consent. Phase~1 interviewees and Phase~2 experts received no compensation; Phase~3 participants received RMB~50.

\begin{table*}[t]
\caption{Overview of the three formative phases. Sample counts describe participation at each phase rather than independent samples: all three Phase~2 experts and one Phase~3 patient had participated in Phase~1.}
\label{tab:formative-overview}
\centering
\small
\begin{tabularx}{\textwidth}{@{}p{0.12\textwidth}p{0.24\textwidth}X X@{}}
\toprule
\textbf{Phase} & \textbf{Participants and activity} & \textbf{Analytic / design role} & \textbf{Primary output} \\
\midrule
Phase~1: Stakeholder Interviews & 12 patients, 12 dental students, and 12 healthcare professionals; 35 synchronous interviews and one written response to the same guide; interviews averaged approximately 45 minutes. & Explore when dental situations become threatening, how they are experienced, what support is desired, and what a VR system should accommodate. Two researchers used a three-round hybrid deductive--inductive analysis. & Three Anxiety Threats (T1--T3), five candidate intervention approaches (S1--S5), and three cross-cutting system considerations (H16--H18). \\
\addlinespace
Phase~2: Expert Participatory Design & Three returning outpatient dentists in one joint session of approximately 30 minutes; card sorting, combination, and workflow modification on an A2 sheet. & Assess clinical plausibility and functional appropriateness and translate the Phase~1 design space into one coherent dental journey. & Unilateral impacted mandibular third-molar extraction under local anaesthesia; preliminary E1--E5; M0--M4; and a preliminary workflow. \\
\addlinespace
Phase~3: Patient Walkthrough Validation & 12 patients; one common approximately 10-minute no-intervention Meta Quest~3 walkthrough followed by an approximately 15-minute interview. & Examine patient-facing plausibility, ordering, comprehensibility, supervised tolerability, and presentation artefacts before the intervention package was implemented. & Event- and transition-level refinements to E1--E5 and the simulation workflow. \\
\bottomrule
\end{tabularx}
\end{table*}

\subsection{Phase 1: Stakeholder Interviews}
\label{sec:formative-phase1}

Phase~1 interviewed 36 stakeholders across three groups---patients, dental students, and healthcare professionals---contributing complementary first-person, training, and clinical perspectives (composition and eligibility in Table~\ref{tab:formative-overview} and Appendix~\ref{app:formative-participants}). Two researchers analyzed the Chinese-language transcripts through a three-round hybrid deductive--inductive procedure in Miro, yielding 18 themes organized around what makes a dental situation threatening, what support is sought, and what the system should accommodate (coding workflow and full inventory in Appendix~\ref{app:phase1-analysis}).

The synthesis produced three Anxiety Threats and five support approaches, summarized with their design translations in Table~\ref{tab:formative-constructs}. \textbf{T1 Procedural Uncertainty}---``The most stressful part is waiting without knowing what the dentist will do next. I keep wondering when it will start and how long it will last'' (P1-P10). \textbf{T2 Sensory and Invasive Threat}---``When I see the needle coming toward me, I tense up before it even touches me'' (P1-P09). \textbf{T3 Expectation Violation}---``After the anaesthetic, or when the drill stops, I expect the difficult part to be over. If I then feel pain or they start again, the anxiety comes back immediately'' (P1-P04). Support needs (S1--S5) ranged from preserving agency (``I would feel safer if I could raise my hand and know the dentist would pause,'' P1-P06) to arousal regulation, procedural guidance, sensory attenuation, and non-directive reassurance. These categories describe recurring patterns rather than diagnoses or prevalence estimates. Phase~1 established the design space---interpreted through the Cognitive Vulnerability Model for threat appraisals---but did not determine the focal procedure, finalize events, or specify module assignments; Gross's Process Model is introduced in Section~\ref{sec:system} to organize the regulatory rationale of the implemented modules.

\begin{table*}[t]
\caption{Phase~1 constructs and their subsequent design translation. Event and module specifications were developed after Phase~1.}
\label{tab:formative-constructs}
\centering
\small
\begin{tabularx}{\textwidth}{@{}>{\raggedright\arraybackslash}p{0.25\textwidth}>{\raggedright\arraybackslash}X>{\raggedright\arraybackslash}p{0.28\textwidth}@{}}
\toprule
\textbf{Construct} & \textbf{Core threat or support need} & \textbf{Later design translation} \\
\midrule
T1 Procedural Uncertainty & Uncertain timing, actions, sensations, or progress. & E1: uncertain waiting. \\
T2 Sensory and Invasive Threat & Threatening instruments, sounds, or bodily sensations. & E2: instrument exposure; E3: drilling-audio onset. \\
T3 Expectation Violation & Events conflicting with established expectations. & E4: unexpected cue after anticipated numbness; E5: continued treatment after apparent completion. \\
\midrule
S1 Agency and Safe Disengagement & Signal discomfort; pause, stop, or reduce stimulation. & M0: pause, resume, and exit. \\
S2 Arousal Self-regulation & Brief breathing guidance or time to regain composure. & M1: arousal self-regulation. \\
S3 Procedure Guidance & Timely explanations of steps, sensations, and progress. & M2: procedure guidance. \\
S4 Sensory Regulation & Attenuate threatening cues while retaining procedural context. & M3: visual and auditory sensory titration. \\
S5 Emotional Support & Acknowledgement, reassurance, and companionship without outcome guarantees. & M4: emotional support. \\
\bottomrule
\end{tabularx}
\end{table*}

\subsection{Phase 2: Expert Participatory Design}
\label{sec:formative-phase2}

Three Phase~1 outpatient dentists returned for a joint participatory-design session using Phase~1-derived Anxiety Event cards, Module cards, and an A2 design sheet (session materials and all six refinements in Appendix~\ref{app:phase2-details}). They recommended unilateral extraction of an impacted mandibular third molar under local anaesthesia---a staged, bounded scenario accommodating all three threat categories with a stable patient viewpoint---and produced five preliminary Anxiety Events (E1--E5) alongside the M0--M4 module framework. Four refinements were especially consequential: pause/stop controls were separated as an always-available M0 foundation; procedural information was tied to the unfolding action rather than front-loaded; sensory reduction was reframed as selective attenuation; and reassurance was reframed as non-directive recognition rather than outcome guarantees. The experts also cautioned against concurrent support streams, informing a low-interference design goal, though the final one-module-per-event assignment was a subsequent research-team decision.

\subsection{Phase 3: Patient Walkthrough Validation}
\label{sec:formative-phase3}

We next recruited 12 patients who had attended a dental visit within the previous six months; one had also participated in Phase~1 and the other 11 had not. Every participant experienced the same preliminary no-intervention prototype before any cohort-level revisions were implemented. After a baseline interview about dental experiences and pressure points, participants completed an approximately 10-minute Meta Quest~3 walkthrough containing the preliminary E1--E5 sequence without M1--M4 or the final system-controlled pause menu; a researcher remained beside each participant and could pause the experience on a raised-left-hand signal. Participants then commented on event plausibility, ordering and transitions, comprehensibility, perceived tolerability and safety, and needed refinements. Appendix~\ref{app:phase3-walkthrough} reports the walkthrough protocol and the consolidated prototype and refinement history.

The Phase~3 interviews were analyzed using the same three-round hybrid deductive--inductive procedure and cross-checking process as Phase~1, but with a narrower design focus: the face plausibility of the preliminary events; coherence and comprehensibility of ordering, transitions, and presentation; perceived tolerability and safety; and workflow revisions. Table~\ref{tab:phase3-refinements} in Appendix~\ref{app:phase3-walkthrough} links this feedback to subsequent refinements; Section~\ref{sec:scenario-design} and Figure~\ref{fig:event-storyboard} describe the implemented events.

No participant discontinued the walkthrough because of discomfort; this result reflects completion under the supervised Phase~3 procedure rather than the absence of discomfort or general safety of later-added features. Phase~3 provides qualitative evidence about face plausibility, comprehensibility, and supervised tolerability of the common no-intervention prototype. It does \emph{not} establish clinical equivalence, therapeutic effectiveness, or validation of M1--M4, the final in-VR controls, or post-cohort revisions. These boundaries are detailed in Appendix~\ref{app:evidence-boundaries}.

\subsection{Cross-Phase Synthesis: Design Goals}
\label{sec:formative-goals}

Across the three phases, stakeholder accounts established the threat/support design space, expert input translated that space into a clinically plausible scenario and module framework, and patient walkthroughs refined how the resulting events and transitions should be presented. The research team synthesized these inputs into four design goals:

\begin{description}
    \item[DG1: Address event-related support needs.]
    Support should respond to the informational, sensory, self-regulatory,
    or interpersonal needs elicited by the current dental event.
    An appropriate response may clarify what is happening, attenuate
    threatening cues, support arousal regulation, or provide reassurance;
    it need not remove the source of the threat.
    \item[DG2: Preserve agency and safe disengagement.] Participants should be able to pause the experience and decide whether to continue or leave without waiting for an event-specific intervention. 
    \item[DG3: Provide proportionate and minimally disruptive support.] Support should respond to the immediate need while preserving the information and continuity required to follow the dental scene. This is addressed through a one-module-per-event constraint and through sensory titration that attenuates rather than removes threatening cues.
    \item[DG4: Make the experience and support comprehensible.] Participants should be able to follow procedural transitions and understand the purpose of the support presented to them, supported by an interaction tutorial, contextual narration by the virtual dentist, and visible state feedback for all controls.
\end{description}

These goals synthesize Phase~1 threats, support needs, and system considerations (T1--T3, S1--S5, H16--H18), Phase~2 expert refinements concerning agency, contextual information, selective attenuation, and non-directive reassurance, and Phase~3 patient feedback on event presentation and transitions. They express design priorities rather than strict engineering specifications; their realization in the implemented system and their evaluation are described in Sections~4 and~5, respectively. The Phase~1 threat and support findings address RQ1, while their translation through Phases~2 and~3 into events, modules, and a coherent workflow provides the formative design basis for RQ2.

\section{System Design and Implementation}
\label{sec:system}

We translated the formative findings into a simulated dental procedure with five Anxiety Events and a predefined intervention package. The following sections describe the scenario, event representations, support functions, and implementation.

\subsection{Scenario Design}
\label{sec:scenario-design}
\label{sec:scenario-selection}
\label{sec:system-events}

Following the three Phase~2 dentists' recommendation, we selected unilateral extraction of an impacted mandibular third molar under local anaesthesia. This bounded scenario brings together anticipated pain, invasive instruments, procedural uncertainty, and sensory cues associated with extraction anxiety~\cite{astramskaite2016extraction,oosterink2008stimuli,earl1994thirdmolar}. Its staged structure allowed the threats identified in Phase~1 to be represented as reproducible events. Expert review informed the scenario and event specification; Phase~3 patients assessed the preliminary sequence's plausibility and comprehensibility.

The scenario provided a common context for visual, auditory, informational, and expectation-related threats. Restricting the sequence to one procedure enabled event order and intervention points to be specified in advance for each condition. The resulting events are representations of anxiety-provoking situations, rather than a simulation of every clinical step or sensation involved in extraction. In particular, the design distinguishes uncertainty before an action from a later violation of an established expectation. These distinctions guided E1's limited advance information and E4--E5's explicit establishment of a preceding expectation.

The participant experiences E1--E5 once, in a fixed order, from a dental-chair viewpoint (Figure~\ref{fig:event-storyboard}). E1, \emph{Fearing an Unknown Dental Procedure}, introduces waiting with limited information before a critical step. E2, \emph{Seeing Dental Instruments}, brings needles, drills, or other invasive instruments into view. E3, \emph{Hearing Drill Sounds}, introduces drilling audio abruptly. E4, \emph{Feeling Unexpected Pain After Numbness}, first establishes an anaesthesia context through the dentist's actions and narration. Controller vibration then symbolizes an unexpected pain-related sensation, reinforced by a peripheral red gradient. This cue does not reproduce oral pain. E5, \emph{Experiencing Unexpected Continuation of Procedure}, uses apparent-completion cues: drilling stops, the instrument withdraws, and a pause follows. The dentist then introduces an additional operation without previously declaring that the procedure had ended. The E4 gradient and revised E5 completion cues were added after the common Phase~3 walkthrough and were not retested by that cohort (Appendix~\ref{app:phase3-walkthrough}).

\subsection{System Design}
\label{sec:vr-system-design}
\label{sec:system-overview}

The system combines the dental environment, participant controls, event-specific support, and a scripted E1--E5 trajectory (Figures~\ref{fig:system-teaser} and~\ref{fig:system-overview}). Four layers organize these functions. The scenario layer presents the environment and events; the intervention layer implements M0--M4. The orchestration layer executes the predefined event--module script. The logging layer records event boundaries, module onset and offset, participant controls, and head-directed orientation. Delivery is event-contingent: recorded physiological and behavioural data do not select or reschedule modules.

\subsubsection{Tutorial Session}
\label{sec:interaction-tutorial}

Before the simulation, participants completed an approximately
5-minute tutorial introducing the interaction sequence.
They practised using the left-hand gesture and pause menu described
under M0 below.

\subsubsection{Simulation Session}
\label{sec:simulation-stage}
\label{sec:event-module-sync}
\label{sec:system-modules}

\begin{figure*}[t]
  \centering
  \includegraphics[width=\textwidth]{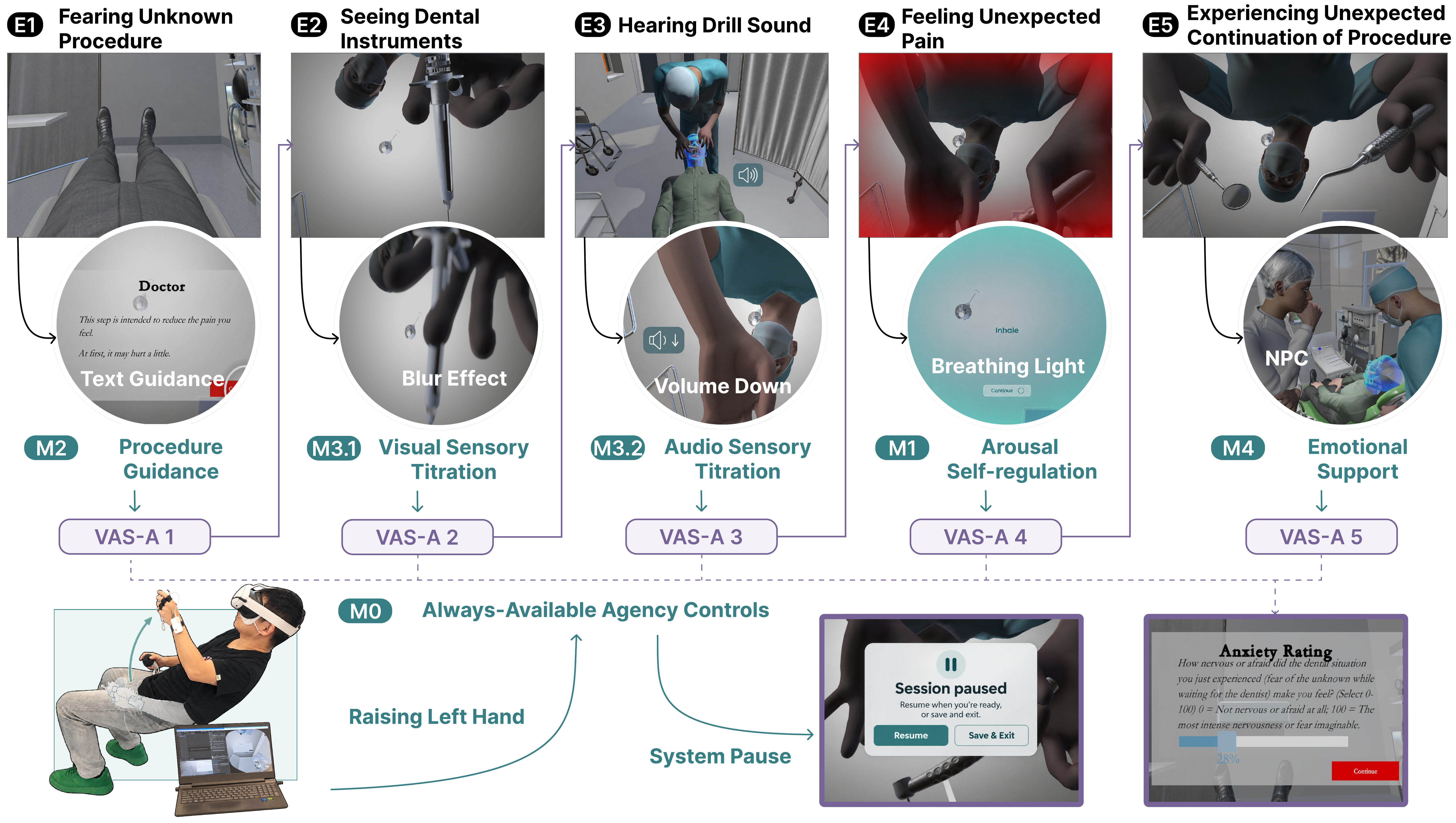}

  \caption{Scripted event--module sequence and interaction
  interfaces of the VR system.
  The upper sequence shows the intervention condition:
  E1--E5 are paired with M2 procedure guidance,
  M3.1 visual sensory titration, M3.2 auditory sensory titration,
  M1 arousal self-regulation, and M4 emotional support,
  respectively, followed by the corresponding event-specific
  VAS-A rating.
  M3.1 and M3.2 are two realizations of M3.
  E4 uses controller vibration and a peripheral red gradient
  as symbolic pain-related cues, rather than reproducing oral pain.
  The lower portion illustrates the left-hand pause gesture,
  the pause menu, and an example VAS-A interface.
  M0 remains available throughout both conditions.
  Control follows the same E1--E5 sequence with M0 but without
  M1--M4 or a companion.
  M2 follows the logged E1 offset; the companion appears only
  during E5 in the intervention condition, after drilling resumes.
  Arrows indicate schematic progression rather than elapsed time.}

  \Description{Five columns run from E1 to E5.
  Rectangular VR scene images show waiting, approaching dental
  instruments, drilling, a scene with a red peripheral gradient,
  and continued dental operation.
  Overlapping circular insets show procedural text, blurred
  instrument detail, sound-reduction annotations, a breathing
  light, and a companion character.
  Each column leads to a VAS-A rating box and then to the next
  event. A dashed connector links the rating boxes to an example
  questionnaire interface at the lower right.
  At the lower left, a photograph shows a seated headset user
  raising the left hand. Arrows connect this gesture to M0 and
  a paused-session menu with Resume and Save \& Exit buttons.}

  \label{fig:system-overview}
  \label{fig:event-storyboard}
  \label{fig:event-module-timeline}
  \label{fig:module-interfaces}
\end{figure*}

Phase~2 established M0--M4 and highlighted competition between concurrent support content (EX5, Appendix~\ref{app:phase2-details}). The research team subsequently assigned one support module to each event, informed by its CVM-related vulnerability and Gross's Process Model~\cite{armfield2006cvm,gross1998emotion}. Figure~\ref{fig:event-module-timeline} illustrates the fixed
event--module assignments, the subsequent VAS-A ratings,
and the independently available M0 controls. The mapping expresses each module's intended regulatory focus without assuming mutually exclusive mechanisms.

The modules serve different roles within the formative design goals. Event-specific assignments implement DG1 by relating support to the immediate threat. Shared pause and withdrawal controls implement DG2's requirement for agency beyond predefined support periods. The one-module-per-event rule and selective sensory attenuation implement DG3's emphasis on proportionate support. The tutorial, contextual narration, and visible control feedback address DG4's requirement for comprehensibility. These relationships explain the implementation choices; the evaluation examines their combined use rather than separately testing whether each design goal was achieved.

\paragraph{M0: Always-Available Agency Controls.}
M0 provides a participant-initiated agency and safety layer
throughout E1--E5, independently of scheduled support delivery.
Raising the left hand pauses the simulation and opens a menu
from which participants can resume or exit. This implementation
follows the Phase~2 recommendation to separate pause and stop
controls from predefined support moments
(EX1, Appendix~\ref{app:phase2-details}). It replaces the
researcher-mediated pause signal used in the preliminary
Phase~3 walkthrough with an application-controlled interaction.
M0 offers no separate participant-operated stimulation-intensity
control.

\paragraph{E1--M2: Procedure Guidance.}
M2 communicates the current stage, next step, purpose, expected duration, and possible sensations. It addresses the prediction gap associated with an unknown procedure through cognitive change. The guidance period follows the E1 boundary, before its VAS-A rating.

This stage-related guidance reflects the experts' distinction between preparatory information and explanations close to the relevant action (EX2). Its content gives meaning to the upcoming procedure while leaving breathing regulation, sensory modification, and companionship to their respective modules.

\paragraph{E2--M3.1 \& E3--M3.2: Sensory Titration.}
M3 modifies threatening sensory input while retaining the underlying event. At E2, M3.1 locally masks or blurs invasive instrument detail after exposure, supporting situation modification and attentional deployment. At E3, M3.2 reduces drilling-sound intensity after onset, supporting situation modification. These visual and auditory realizations belong to one module and are delivered separately at their assigned events.

Both realizations follow the expert refinement from global removal toward selective attenuation (EX3). Instrument detail can be reduced while the broader dental scene remains available; drilling intensity can be reduced without eliminating all auditory context. The design therefore retains information needed to follow the procedure.

\paragraph{E4--M1: Arousal Self-regulation.}
M1 presents a breathing light and textual breathing cues, with a brief guided-breathing pause where specified by the script. It supports response modulation after the unexpected pain-related cue, without assigning a clinical cause to that cue.

\paragraph{E5--M4: Emotional Support.}
M4 provides spoken companionship through a non-directive companion NPC, complementing the intrapersonal processes in the Process Model~\cite{zaki2013interpersonal}. After the apparent-completion interval, the dentist explains the additional operation and drilling resumes. The NPC enters after a scripted interval of continued drilling and verbally offers to accompany the participant. It appears only during E5 in the intervention condition and adds no procedural instructions. VAS-A5 follows the support period.

\subsection{Implementation Details}
\label{sec:implementation}

The application ran on Meta Quest~3 and used Unity~6000.3.12f1, URP~17.3.0, Meta XR Core SDK~205.0.0, and Unity OpenXR Plugin~1.17.1. Physically based materials, baked lighting, a base render scale of 0.8, and $4\times$ multisample anti-aliasing supported mobile rendering. Interaction used tracked Meta Touch controllers rather than optical hand tracking. Unity Animator and NodeCanvas Dialogue Trees coordinated predefined character animations, dialogue, and feedback. A custom manager provided three-dimensional audio with distance attenuation and an AudioListener at the participant's XR camera. Controller vibration supplied E4's symbolic haptic cue.

A scripted controller managed the tutorial, events, and module periods. It detected the pause gesture when left-controller displacement exceeded a predefined threshold within a 1-second interval. The application logged event and module onset/offset, control actions, and head-directed orientation using application-relative time and absolute UTC timestamps. Records were exported locally as JSON, Excel, and SQL files.

The logged variable \emph{Head Gaze} represents head-directed orientation, not ocular gaze. Raycasts within a cone around the headset's forward direction retained each ray's nearest collider intersection. Targets were ranked by angular offset and then hit distance, yielding one primary target per sampling interval. Appendix~\ref{app:head-orientation} specifies sampling, target availability, episode qualification, and temporal aggregation.

\section{User Evaluation}
\label{sec:user-study}

We conducted a two-arm randomized between-participant evaluation to address RQ3, comparing the complete intervention package with the same dental-event sequence without M1--M4.

\subsection{Study Design and Participants}
\label{sec:study-design}

We recruited 24 adults through a local hospital and a university campus. Eligible participants were at least 18 years old and either had attended a dental visit within six months or reported dental fear or anxiety. They were not required to be awaiting third-molar extraction. Researchers asked participants to confirm the absence of known cardiac disease or other physiological conditions making participation unsuitable; this was not standardized medical screening.

A researcher assigned numeric identifiers, randomly shuffled them, and divided participants equally between intervention and control ($n=12$ each). Assignment did not use MDAS or demographic stratification. Participants could not be blinded to intervention delivery; analysts conducting between-group quantitative comparisons were blinded to assignment. Table~\ref{tab:study-descriptives} reports participant characteristics, and Section~\ref{sec:analysis-coverage} reports outcome-specific coverage.

The resulting groups differed in age and gender composition. Five intervention participants and 11 control participants were aged 18--24; the respective men/women counts were 7/5 and 4/8. Mean MDAS scores were 15.58 (SD~3.29) and 13.25 (SD~4.37), respectively. Baseline current VAS-A averaged 10.75 (SD~11.75) and 14.24 (SD~17.34). Six intervention participants and four control participants reported prior impacted-third-molar extraction. Age was recorded in bands, so an exact mean age was unavailable. These characteristics describe this sample; the adjusted primary model included MDAS and baseline current VAS-A but not age or gender.

The controlled-study protocol was approved by the Ethics Committee of Shenzhen Stomatology Hospital (Pingshan) of Southern Medical University (approval no.~202613A). Researchers explained data collection and experimental risks before obtaining informed consent. Participants received RMB~50 after completing the experiment.

\subsection{Apparatus and Setting}
\label{sec:study-apparatus}

Participants used a Meta Quest~3 headset with tracked controllers.
A Shimmer3R GSR unit recorded EDA and photoplethysmography (PPG)
through the manufacturer's SDK with default acquisition settings
(Figure~\ref{fig:user-study-procedure}, B.1).
Participants wore disposable VR face masks beneath the headset
for hygiene (Figure~\ref{fig:user-study-procedure}, B.2). HR was estimated from optical pulse signals rather than ECG. Physiological coverage checks used a sampling rate of 128~Hz (Appendix~\ref{app:study-analysis-details}). The unit was strapped to the left wrist. EDA electrodes contacted the palmar middle phalanges of the left index and middle fingers; the PPG sensor contacted the left index fingertip. Data were stored locally for processing.

Sessions took place in a quiet, otherwise unused room.
During the VR simulation, participants reclined in chairs to
approximate a dental-chair posture and align their physical
position with the simulated patient viewpoint
(Figure~\ref{fig:user-study-procedure}, C.1--C.3).
Three researchers respectively provided instructions and
monitored safety, monitored physiological signals, and supervised
the application's scripted progression.

\subsection{Procedure}
\label{sec:study-procedure}

\begin{figure*}[t]
  \centering
  \includegraphics[width=\textwidth]{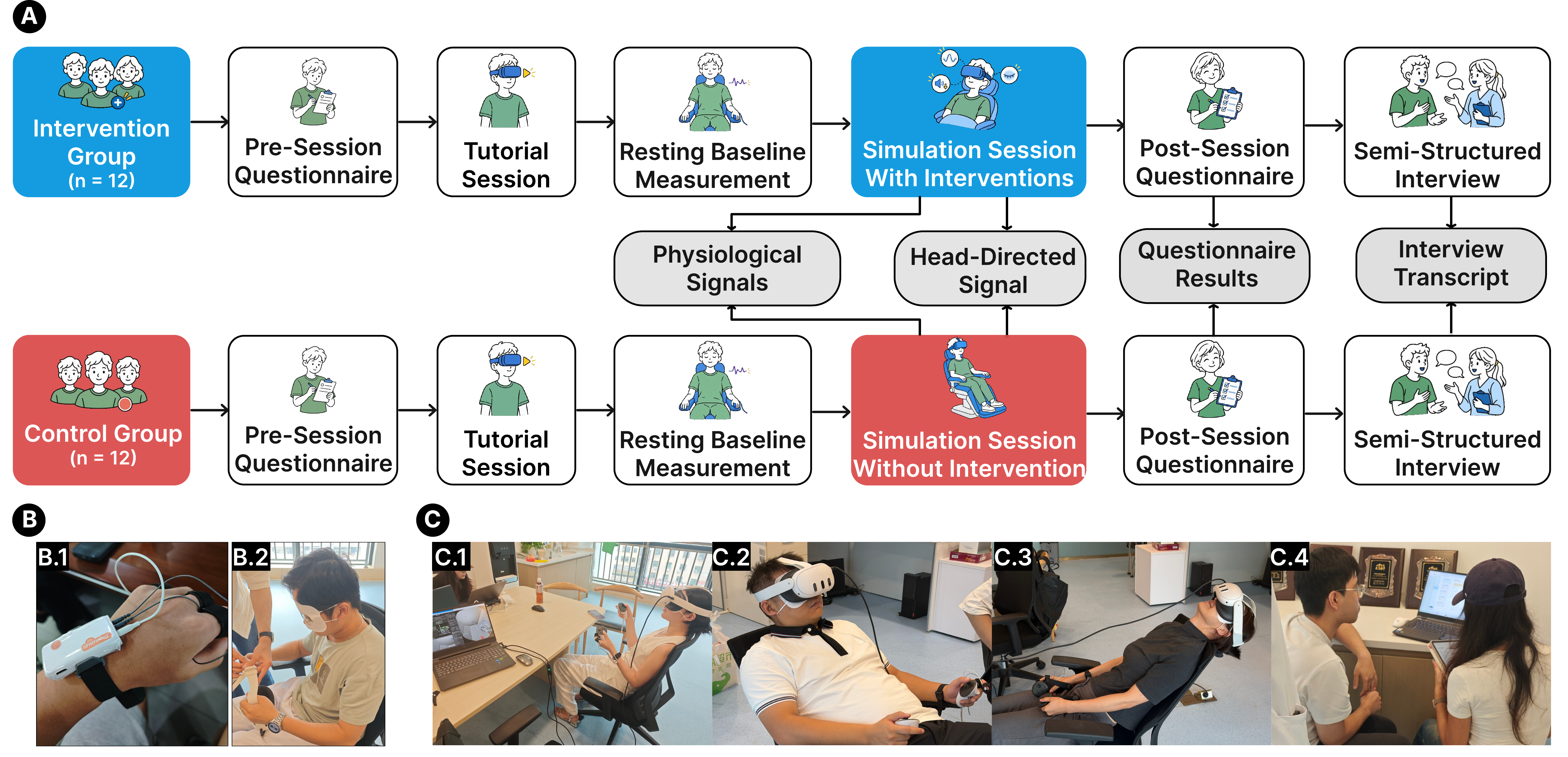}
  \caption{User evaluation procedure and experimental setup.
(A) Workflow of the randomized between-participant evaluation
($N=24$, $n=12$ per condition). Both conditions shared the tutorial,
resting-baseline procedure, E1--E5 sequence, and M0 controls.
Simulation runs lasted approximately 15 minutes with M1--M4 and
10 minutes without them. Event-specific VAS-A ratings, continuous
physiological recordings, and headset-based orientation logs
were collected during the simulation.
(B) Pre-session preparation: (B.1) the Shimmer3R GSR unit for
electrodermal activity and photoplethysmography recording, and
(B.2) a disposable VR face mask for hygiene.
(C.1--C.3) Study photographs showing participants using Meta
Quest~3 while reclining to approximate the posture and viewpoint
of a patient in a dental chair.
(C.4) A post-session semi-structured interview.}

\Description{A composite figure places a flowchart above two
groups of photographs. In panel A, a blue upper row represents
the intervention condition and a red lower row represents the
control condition, each with 12 participants. Left-to-right
arrows connect pre-session questionnaires, a tutorial, resting
baseline measurement, the assigned simulation, post-session
questionnaires, and a semi-structured interview. Central boxes
identify physiological signals, head-directed orientation,
questionnaire responses, and interview transcripts.
Below the flowchart, B.1 shows a small recorder secured to a
participant's wrist, with leads extending toward the fingers.
B.2 shows a participant wearing a white disposable mask around
the eyes before donning the VR headset.
C.1--C.3 show participants wearing VR headsets and holding
controllers while reclining in chairs beside the study equipment.
C.4 shows two people seated at a desk in conversation.}
  \label{fig:user-study-procedure}
\end{figure*}

Figure~\ref{fig:user-study-procedure}A summarizes the study workflow.
Both groups completed background and safety-related questions,
the Chinese MDAS, and baseline current VAS-A. A shared tutorial lasting approximately 5 minutes preceded approximately 2--3 minutes of eyes-closed resting physiological recording in a black scene, with signal quality checked before proceeding. The subjective baseline therefore preceded the tutorial, whereas the physiological reference followed it.

Participants experienced E1--E5 once in a fixed order, using the same
equipment, general instructions, and M0 controls. The intervention workflow
included M1--M4 and their associated presentation and interaction periods;
the control omitted these modules and contained no duration-matching waits.
The resulting runs lasted approximately 15 and 10 minutes, respectively,
excluding the tutorial, resting baseline, questionnaires, and interviews.
This comparison therefore assessed the complete support package as
delivered, including its additional time. Intervention participants rated
each event after its assigned support; control participants rated the
corresponding event without M1--M4. Physiological signals were recorded
continuously throughout the run.

Both groups then completed current VAS-A and VRSQ; intervention participants additionally completed module and system evaluations. Post-session semi-structured interviews lasted approximately 5--10 minutes, with available recordings documented in Section~\ref{sec:analysis-plan}.

\subsection{Data Collection}
\label{sec:study-measures}
\label{sec:data-collection}

Each session produced subjective ratings, physiological recordings, headset-based orientation logs, and a post-session interview, collected as follows.

\paragraph{Subjective measures.}
Background questions documented dental and VR experience, motion-sickness susceptibility, and recent medication, caffeine, and exercise. The primary outcome was event-specific VAS-A, rating nervousness or fear about the dental situation just experienced from 0 (none) to 100 (strongest imaginable). The five ratings were repeated single-item observations; pre- and post-session current VAS-A instead assessed anxiety at that moment. Broader dental anxiety was assessed using the five-item Modified Dental Anxiety Scale (MDAS)~\cite{humphris1995mdas}, administered in the Chinese version evaluated by Yuan et al.~\cite{yuan2008mdas}. Item scores were summed (range: 5--25) and used as a continuous covariate. The nine-item VRSQ assessed VR-sickness symptoms separately from reported adverse events~\cite{kim2018vrsq}.

\paragraph{Intervention evaluations.}
Intervention participants completed four AIM and four IAM items per module (M0--M4), assessing acceptability and appropriateness on 1--5 scales~\cite{weiner2017implementation}. M3 combined its visual and auditory implementations. Five TFA-informed, researcher-developed diagnostic items separately assessed timing, duration, minimal disruption, agency, and delivery transparency~\cite{sekhon2017acceptability}. Appendix~\ref{app:study-questionnaires} provides the wording, scoring, and adaptations, which were not independently validated.

\paragraph{Continuous records and interviews.}
HR and EDA characterized arousal; logs recorded headset-based orientation, event/module timing, and control use. Post-session semi-structured interviews in Chinese explored the VR experience, perceived support, disruption, and improvements; usable recordings were transcribed for thematic analysis.

\subsection{Analysis}
\label{sec:analysis-plan}

Participants were the independent sampling units, with repeated outcomes nested within participants. We prioritized the overall anxiety contrast; event variation and complementary outcomes provided exploratory detail. Appendix~\ref{app:study-analysis-details} documents software versions, approximation rules, diagnostics, distributional observations, and a sensitivity model omitting both covariates.

\paragraph{Event-specific anxiety.}
We fitted a restricted-maximum-likelihood mixed-effects model with categorical Event, Group, their interaction, and a participant-specific random intercept:
\[
\mathrm{VAS\mbox{-}A} \sim \mathrm{Group} \times \mathrm{Event}
+ \mathrm{BaselineCurrentVAS\mbox{-}A}
+ \mathrm{MDAS}
+ (1 \mid \mathrm{Participant}).
\]
Baseline current VAS-A and MDAS were centred across the 24 participants. We estimated intervention-minus-control differences for each event and an equally weighted overall difference across E1--E5, expressed in VAS-A points (negative values indicate lower intervention ratings). Percentile 95\% confidence intervals used 2,000 within-condition bootstrap resamples of whole participants, preserving repeated observations; these intervals describe pointwise uncertainty rather than simultaneous coverage. Model-based $t$ tests and a marginal Group~$\times$~Event $F$ test provided auxiliary inference, with Holm correction covering the five event-specific $p$-values. The overall contrast and interaction addressed different questions and were reported separately. The unadjusted sensitivity model retained the Group~$\times$~Event interaction and participant random intercept; it assessed dependence on covariate adjustment rather than resolving all distributional assumptions.

\paragraph{Physiological outcomes.}
Each outcome used the 20~s immediately preceding its corresponding VAS-A prompt, referenced to the final 60~s of rest. HR was the window mean minus the resting mean, in bpm. EDA was the natural logarithm of the window-to-resting mean ratio, using total EDA rather than isolated phasic responses. Both outcome and reference windows required at least 80\% valid sample coverage; invalid windows were excluded without imputation. Group contrasts were unadjusted differences in group means, with percentile 95\% confidence intervals from 20,000 bootstrap resamples and two-sided $p$-values from 50,000 participant-level group-label permutations of the absolute Welch statistic. Holm correction was applied separately within each outcome's five-event family. Overall contrasts required five valid windows per participant and therefore describe outcome-specific complete-case subsets. All contrasts concern pre-VAS windows that occupied different elapsed-time positions and intervention contexts across conditions. Bootstrap intervals were computed separately from the permutation tests; interval inclusion of zero and permutation thresholds can therefore yield different conclusions. Appendix~\ref{app:study-analysis-details} reports coverage denominators, exclusion criteria, and event-specific sample counts.

\paragraph{Questionnaire summaries.}
Module-specific AIM and IAM scores were four-item means, summarized descriptively alongside the five diagnostic-item distributions. We conducted no between-module tests and did not score M3.1 and M3.2 separately. Post-session current VAS-A and VRSQ were secondary descriptive outcomes.

\paragraph{Orientation and pauses.}
For each participant, we divided qualified primary-hit time by declared target-available time across E1--E5, summing both durations before calculating the percentage. Shared targets were the dentist's face, hands, and body, the anaesthetic instrument, and the drill. NPC orientation was summarized separately for intervention participants, using availability from M4 onset to E5 offset; control's absent NPC was not coded as zero orientation. Pause summaries described participant and episode counts, duration, and resumption, without treating pauses as anxiety scores or evidence of M0's causal effect. We reported medians and interquartile ranges without significance tests. Appendix~\ref{app:head-orientation} defines the head-pose sampling specification, episode qualification rules, event boundaries, target availability, and tracking limitations.

\paragraph{Interview analysis.}
We analyzed 19 transcribed interviews: all 12 intervention interviews and seven control interviews. Five hospital-based control sessions lacked usable transcripts because of scheduling constraints or environmental noise; retrospective researcher notes for three of these were excluded. Two researchers conducted two rounds of thematic grouping, first organizing comments about the VR experience and interventions, then consolidating related groups into the reported themes. This focused analysis was distinct from the formative study's three-round process. Quotations are English translations of Chinese transcripts.

\section{Results}
\label{sec:results}
\label{sec:analysis-coverage}

The analysis included 24 participant records, with 12 per condition and five event-specific ratings per record, yielding 120 repeated observations. Age and gender distributions differed between conditions; Table~\ref{tab:study-descriptives} reports these characteristics alongside dental history and baseline scores. Of 120 physiological windows per outcome, 115 HR windows and 119 EDA windows met the coverage criterion. Complete-five-event comparisons included 8 intervention and 11 control records for HR, and 11 intervention and 12 control records for EDA. All 24 records contributed to the supplementary E1--E5 orientation summaries. The intervention questionnaire contained 12 responses per diagnostic item and 12 AIM and IAM scores per module. Qualitative analysis covered 19 transcripts: 12 intervention and seven control interviews. Section~\ref{sec:analysis-plan} accounts for the five control sessions without usable transcripts.

No adverse events were reported during the study sessions. VRSQ total means were 5.56 (SD 4.58) in the intervention condition and 11.46 (SD 12.39) in the control condition. The questionnaire distributions included nonzero symptom scores in both conditions (Table~\ref{tab:study-descriptives}).

\subsection{Event-specific Dental Anxiety}
\label{sec:subjective-results}

\begin{figure}[t]
  \centering
  \includegraphics[width=\linewidth]{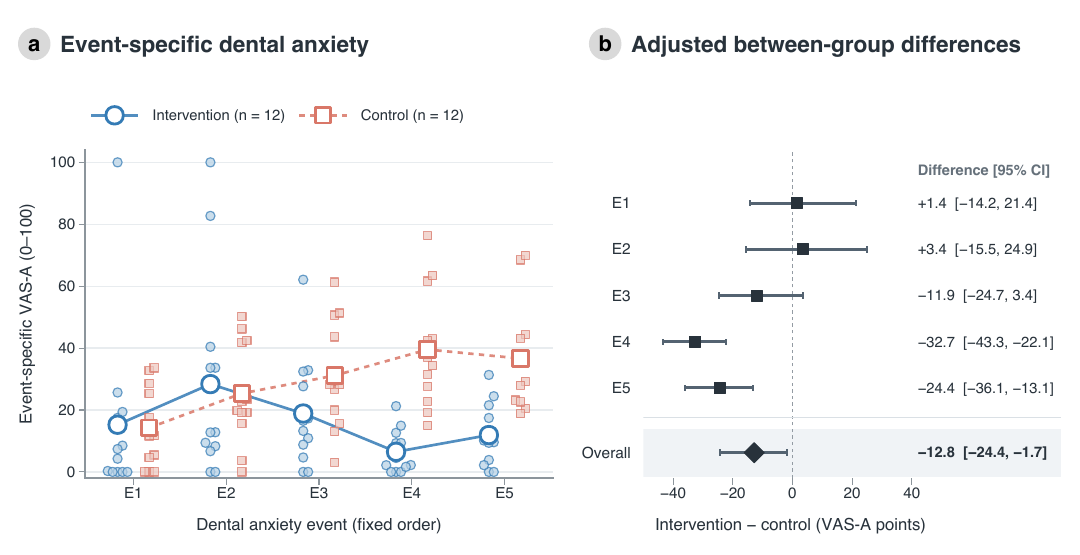}
  \caption{(a) Small marks show individual VAS-A values and large marks show unadjusted group means: blue circles and solid lines denote intervention, and coral squares and dashed lines denote control. Lines indicate the fixed event order. Each of 24 participant records contributes five ratings ($n=12$ per condition). (b) Intervention-minus-control contrasts from the mixed-effects model adjusting for baseline current VAS-A and MDAS, with pointwise 95\% participant-cluster bootstrap percentile confidence intervals from 2,000 resamples. ``Overall'' equally weights the five events. Negative values indicate lower intervention-condition ratings.}
  \label{fig:event-specific-anxiety}
  \Description{Two side-by-side panels are identified by lower-case letters on gray circles. Panel a shows individual anxiety ratings and group means across E1 through E5 on a zero-to-100 scale. Raw intervention means are higher at E1 and E2 and lower at E3 through E5, with substantial individual variation. Panel b is a forest plot of adjusted group differences, with squares for event-specific estimates, horizontal confidence intervals, and a vertical dashed line at zero. A diamond on a shaded row represents the overall estimate of minus 12.83 points, with an interval from minus 24.42 to minus 1.70. E1 and E2 estimates are positive and E3 through E5 estimates are negative. The E4, E5, and overall intervals lie entirely below zero; these are pointwise rather than simultaneous multiple-comparison intervals.}
\end{figure}

The overall event-specific VAS-A contrast favoured the intervention condition (Figure~\ref{fig:event-specific-anxiety}). Averaged equally across E1--E5, the adjusted intervention-minus-control contrast was $-12.83$ points (95\% CI $[-24.42,-1.70]$; auxiliary $p=.040$). The direction and magnitude of the contrast varied across the event sequence.

The Group~$\times$~Event interaction was supported by the approximate marginal test ($F(4,88)=6.68$, $p<.001$). Adjusted contrasts were positive at E1 and E2 and negative at E3--E5. E4 had the largest negative estimate ($-32.66$ points, 95\% CI $[-43.31,-22.12]$; $p_{\mathrm{Holm}}=.0029$), followed by E5 ($-24.36$ points, 95\% CI $[-36.14,-13.15]$; $p_{\mathrm{Holm}}=.0253$). These two event-specific comparisons met the .05 threshold after Holm correction; E1--E3 did not. Table~\ref{tab:vas-contrasts} reports all estimates and auxiliary tests.

As a secondary descriptive outcome, post-session current VAS-A averaged 10.17 (SD 10.45) in the intervention condition and 18.08 (SD 8.75) in the control condition. Omitting baseline current VAS-A and MDAS from the event-specific model yielded a similar overall contrast of $-13.21$ points (95\% CI $[-23.74,-2.34]$). The unadjusted analysis also retained the E4 and E5 comparisons after Holm correction (Appendix~\ref{app:study-analysis-details}).

\subsection{Physiological Arousal}
\label{sec:physiological-results}

\begin{figure*}[t]
  \centering
  \includegraphics[width=\textwidth]{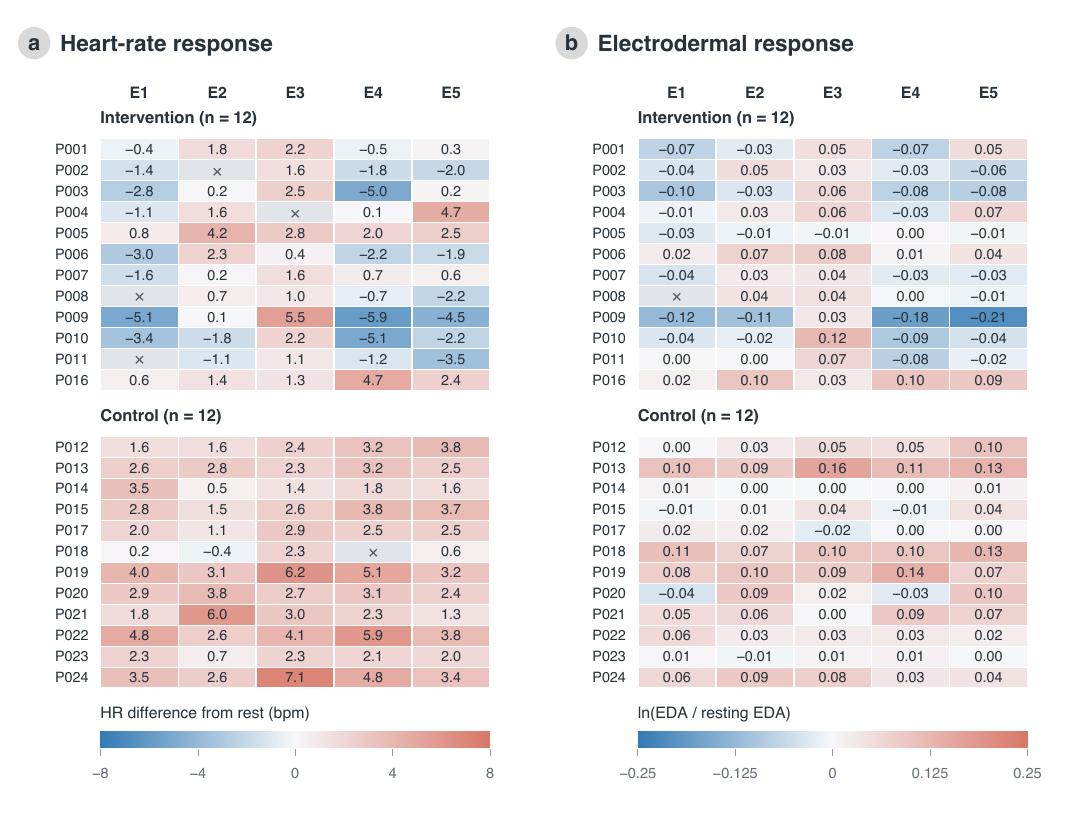}
  \caption{Rows represent participant records grouped by condition ($n=12$ each), with identical participant order across panels. (a) Mean HR in the 20~s before each VAS-A prompt minus its mean in the final 60~s of rest, in bpm. (b) Natural logarithm of the ratio of the corresponding total-EDA means. Blue and coral indicate values below and above rest; condition membership is shown by the row groups. Color scales are symmetric around zero within each outcome. Gray cells marked $\times$ indicate invalid windows. Both the outcome window and resting reference require at least 80\% valid sample coverage: 115/120 HR and 119/120 EDA windows meet this criterion. No missing values are imputed, and no row standardization or clustering is used. Labels are rounded to one decimal for HR and two for EDA; colors use unrounded values.}
  \label{fig:physiological-profiles}
  \Description{Two side-by-side heatmaps have circled panel labels a and b, titled Heart-rate response and Electrodermal response. Each displays five event columns and 24 participant rows, with an upper intervention block and a lower control block. Participant identifiers are ordered identically in both panels. Valid cells contain numerical values and colors; color keys span minus 8 to plus 8 bpm for HR and minus 0.25 to plus 0.25 for EDA, with a pale zero midpoint. Invalid HR cells are P002 at E2, P004 at E3, P008 at E1, P011 at E1, and P018 at E4. The invalid EDA cell is P008 at E1. These cells are gray and marked with a cross.}
\end{figure*}

Rest-referenced pre-VAS HR and EDA values were lower overall in the intervention condition (Figure~\ref{fig:physiological-profiles}). The complete-case unadjusted HR contrast was $-3.01$~bpm (95\% CI $[-4.23,-1.75]$), based on 8 intervention and 11 control records. The corresponding EDA log-ratio contrast was $-0.055$ (95\% CI $[-0.091,-0.021]$), based on 11 intervention and 12 control records.

Event-specific HR contrasts were negative at all five events. EDA contrasts were negative at E1, E2, E4, and E5, and close to zero at E3. Within each outcome's five-comparison family, E1, E4, and E5 met the .05 threshold after Holm correction; E2 and E3 did not (Table~\ref{tab:physiology-contrasts}). At E1, both physiological outcomes met the corrected threshold despite a positive, nonsignificant subjective contrast. The heatmaps show participant-level variation and invalid windows; the supplementary table reports contrasts and uncertainty.

\subsection{Head-Directed Orientation and Pause Records}
\label{sec:head-orientation-results}

Head-directed orientation profiles differed descriptively across the five shared scene regions (Figure~\ref{fig:head-orientation-supplement}). Across E1--E5, the median qualified orientation percentage toward the dentist's face was 21.29\% in the intervention condition and 32.21\% in control. For the drill, the corresponding medians were 14.62\% and 10.23\%. These percentages use region-specific available time, not total session duration. Appendix~\ref{app:head-orientation} reports all five regions, individual values, processing details, and the intervention-only NPC summary.

The pause records contained seven pause--resume episodes across five intervention records, totalling 27.50~s. Four episodes occurred during E1, with one each during E2, E3, and E4; none occurred during E5. The median episode duration was 3.63~s (range: 2.50--5.63~s), and all seven episodes ended in resumption. The records contained no control-condition pause entries; this absence does not establish zero use or support a comparison of usage rates. Appendix~\ref{app:head-orientation} reports the participant identifiers and summarizes the event distribution and pause durations.

\subsection{System and Module Evaluations}
\label{sec:system-evaluation-results}

\begin{figure*}[!t]
  \centering
  \includegraphics[width=\textwidth]{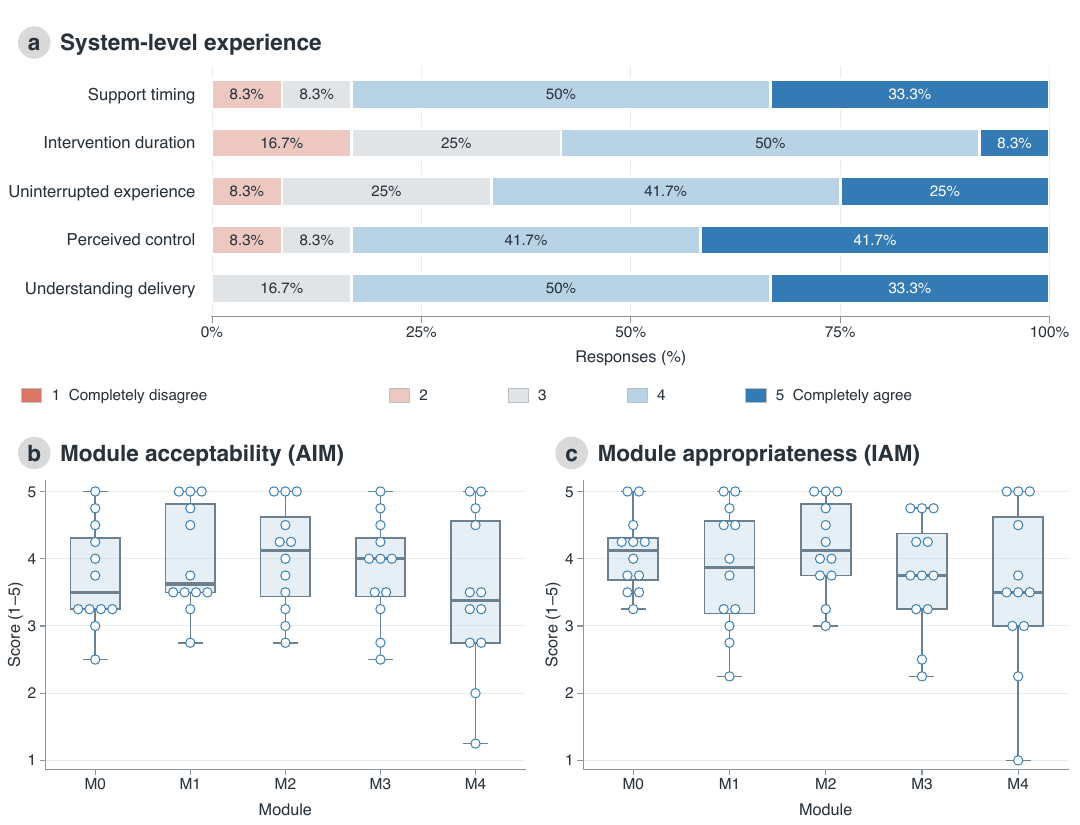}
  \caption{(a) Separate response distributions for five researcher-developed diagnostic items, with 12 responses per item on the original 1--5 scales; 1 indicates completely disagree and 5 completely agree. Segment labels give percentages, which may not sum to 100 after rounding. (b--c) Module acceptability (AIM) and appropriateness (IAM), with each point showing one participant record's four-item mean. Boxes show the median and interquartile range; whiskers reach the most extreme observations within 1.5 interquartile ranges. All points are shown, and the same 12 records contribute to every module. M0 denotes agency and safety; M1, arousal self-regulation; M2, procedure guidance; M3, sensory titration combining its visual and auditory implementations; and M4, emotional support.}
  \label{fig:system-module-evaluations}
  \Description{Three panels are identified by lower-case letters on gray circles. Panel a spans the top and contains five stacked horizontal bars for timing, duration, uninterrupted experience, perceived control, and understanding of delivery. Each bar represents 12 responses ordered from scores 1 through 5, with percentage labels and a shared response-color legend. Scores 4 and 5 comprise 10, 7, 8, 10, and 10 responses, respectively. Panels b and c sit side by side below and show individual AIM and IAM four-item means with box plots on 1-to-5 axes for M0 through M4. Points at the same value are separated horizontally without changing their scores.}
\end{figure*}

Timing, perceived control, and understanding intervention delivery each received scores of 4 or 5 in 10 of 12 responses (Figure~\ref{fig:system-module-evaluations}). The corresponding counts were 7/12 for duration and 8/12 for minimal disruption. Ratings of duration and minimal disruption were therefore less favourable than ratings of timing.

Module-specific mean AIM scores ranged from 3.46 to 4.02 and mean IAM scores from 3.58 to 4.19 on their 1--5 scales. Figure~\ref{fig:system-module-evaluations} shows individual four-item means alongside medians and interquartile ranges; Table~\ref{tab:module-descriptives} provides the numerical summaries. M4 had the lowest descriptive mean on both measures (AIM 3.46; IAM 3.58). These ratings concern acceptability and appropriateness, with visual and auditory sensory titration combined for M3; no between-module tests were conducted. The interviews describe how participants experienced the support.

\subsection{Post-session Interview Findings}
\label{sec:post-session-interviews}

Analysis of 19 transcripts (12 intervention, 7 control) yielded three themes concerning procedural credibility and how breathing guidance (M1) and companionship (M4) may modulate anxiety. Quotations were translated from Chinese and retain anonymized participant identifiers.

\subsubsection{Sensory Realism and Procedural Credibility}

Participants distinguished environmental immersion from procedural credibility, as P6 explained.

\begin{quote}
\emph{``This VR system felt immersive, but I did not feel that the dentist was working on my mouth, so it did not really feel like I was having dental surgery.''} (P6)
\end{quote}

Despite convincing surroundings, absent intraoral tactile and proprioceptive feedback weakened procedural credibility. The convincing room and unconvincing clinical action created a disconnect in P6's experience. Environmental immersion alone may therefore be insufficient without feedback at the treatment site.

P8 contrasted convincing sound with the anticipated effect of another sensory channel.

\begin{quote}
\emph{``The sound was very realistic, but I think the smell of dental disinfectant would make me more anxious.''} (P8)
\end{quote}

Adding smell or intraoral sensation could increase both credibility and anxiety. For gradual exposure, greater sensory realism may therefore intensify the distress that support seeks to mitigate.

Together, these accounts indicate uneven fidelity across modalities, particularly at the treatment site. Participants also requested room transitions, additional dentist speech, and more natural characters. Future iterations should prioritize channels closely tied to dental procedures while examining their effects on anxiety.

\subsubsection{Engagement with Breathing Guidance (M1)}

Participants valued breathing regulation but differed in engagement with its implementation. P6 connected nasal breathing to a practical treatment constraint.

\begin{quote}
\emph{``Breathing regulation is very useful. Sometimes, when lying in a dental chair, saliva means I can only breathe through my nose. Practising this is quite helpful.''} (P6)
\end{quote}

For P6, the exercise addressed a physical demand encountered during actual dental treatment. This account suggests that framing breathing as rehearsal for procedural demands may make it more meaningful than presenting it solely as relaxation.

However, P3 reported a competing attentional demand that disrupted engagement.

\begin{quote}
\emph{``When the breathing light appeared, I kept pressing Continue. My attention was drawn there, so I may not have followed the system's prompts.''} (P3)
\end{quote}

The continuation control competed with attention needed to follow the rhythm. Providing guidance therefore did not ensure that participants could engage with it.

Perceived value and effective engagement were distinct, as these contrasting accounts illustrate. One participant proposed an expanding and contracting circle to clarify the rhythm. Guidance must remain followable without requiring attention to competing interface elements. Separating continuation controls from guidance spatially and temporally could reduce competition.

\subsubsection{Companionship and Perceived Reassurance (M4)}

Participants attributed two social functions to the companion beyond visual presence.

The first concerned emotional support, with P1 describing reduced isolation.

\begin{quote}
\emph{``I think the companion NPC works best, because having a friend there makes me feel that I am not alone.''} (P1)
\end{quote}

P1 valued shared presence rather than instructions or clinical information. Another agent's presence shifted the experience from solitary vulnerability toward a sense of enduring treatment together. This interpretation aligns with social presence theory linking another agent's presence to affective responses~\cite{Skalski28092007}.

The second concerned instrumental reassurance, with P8 describing perceived oversight.

\begin{quote}
\emph{``Companionship is helpful because I am afraid of mistakes during dental procedures. Having another person there feels like having someone to keep an eye on things.''} (P8)
\end{quote}

P8 perceived the companion as watching a procedure patients cannot monitor themselves. Apparent vigilance may reduce the perceived likelihood that procedural errors remain undetected. This perceived advocacy resembles reassurance associated with accompaniment during medical appointments~\cite{baron1993emotional}.

Another participant suggested a familiar friend, raising relational closeness as a design consideration. Familiarity might strengthen emotional support, whereas perceived attentiveness might support vigilance.

These accounts distinguish shared presence from delegated vigilance as possible sources of reassurance. Companion identity and behaviour may activate these functions to different degrees. Future work could examine how matching companionship to participants' dominant anxiety concerns affects its value.

\section{Discussion}
\label{sec:discussion}

This study connects formative dental-anxiety findings with reproducible VR events and support modules. The results reveal tensions among credibility, agency, burden, and reassurance. CBT and exposure research clarify the distinction between immediate support and learning beyond an encounter.

\subsection{Connecting Procedural Appraisals with Cognitive Work}
\label{sec:discussion-event-specific}
\label{sec:discussion-translation}

Procedural uncertainty, sensory threat, and expectation violation involve interpretations as well as cue intensity. E4 and E5 require anticipated numbness or completion before conflicting cues. CVM identifies vulnerability appraisals, while the Process Model distinguishes regulatory actions~\cite{armfield2006cvm,armfield2008dental,gross1998emotion}. CBT additionally asks what participants understand or do differently after support~\cite{beck2021basics}.

M2 explains procedural purpose, sensations, and duration without eliciting or revisiting predictions. Receiving an explanation is therefore distinct from practising a cognitive skill. Reappraisal research highlights difficulties identifying thoughts and generating alternatives during strong emotion~\cite{kitson2024reappraisal}; thought-recording systems make this work explicit~\cite{burger2022thoughtrecord}. A preparatory exercise could elicit concerns about interrupting treatment, rehearse interruption, and invite reflection, connecting information to a testable belief.

Detailed reflection during drilling could compete with comprehension and communication. Coping statements before an event, concise support during it, and reflection at an agreed pause or afterward could distribute cognitive work. The activity would retain its connection to the threatening moment without requiring simultaneous reflection. This extends distinctions between learning and applying regulatory skills~\cite{slovak2023emotion}, but requires comparison with information-only guidance to assess comprehension and later coping.

The largest subjective contrasts occurred at E4 and E5; earlier contrasts did not meet the corrected threshold. This variation motivates examining how participants interpret support after expectations are disrupted. Fixed event--module assignments and sequence positions prevent attributing effects to breathing or companionship individually. Future studies could assess appraisals before and after each event--support configuration and examine the activity it enables, rather than infer cognitive change from anxiety alone.

\subsection{Supporting Engagement and Learning across Encounters}
\label{sec:discussion-exposure-learning}

Immediate anxiety reduction leaves subsequent approach to dental care unresolved. Gujjar et al.'s dental-phobia trial illustrates broader evaluation through follow-up anxiety, avoidance, and treatment acceptance~\cite{gujjar2019dental}. Our findings concern the supported encounter rather than willingness or ability to approach later treatment.

M3 reduces selected sensory cues while retaining the dental context. Exposure-oriented evaluation should distinguish temporary support for engagement from believing coping requires obscured threats. Mixed findings on safety behaviours support examining these different learning possibilities empirically~\cite{blakey2019safety}. A supervised extension could compare stable and gradually reduced optional support, retaining withdrawal controls and assessing approach to a subsequent scene with less assistance. Current settings were neither personalized nor progressively faded during the simulation.

E4 and E5 violate expectations to elicit anxiety; therapeutic expectancy violation instead tests feared predictions against less threatening outcomes~\cite{craske2014inhibitory}. Our later-event anxiety contrasts do not establish that therapeutic learning process. Future exercises could test whether interruption requests are acknowledged without promising painless, predictable care. Participants could then reflect on the observed response in relation to their original prediction about control. Preparatory explanations should follow the learning goal because they may alter the prediction being tested.

Breathing could support voluntary engagement during distress, followed by practice without cues. Success would include independent skill use or re-entry, rather than requiring low arousal before continuing. Neither respiratory performance nor transfer was assessed, so E4 cannot establish skill acquisition. Immediate comfort and longer-term practice therefore require separate assessment against their respective goals.

\subsection{Preserving Agency while Reducing Support Burden}
\label{sec:discussion-agency}

Scheduled modules coexist with participant-initiated interruption through M0's left-hand gesture and pause menu, offering continuation or exit. Interruption remains independent of timing or inferred distress; shared availability prevents isolating M0's anxiety effect.

Safety-behaviour concerns do not justify removing consent, withdrawal, or communication controls. Availability, actual use, and beliefs about what a control accomplishes remain distinct questions. Recorded pauses ended in resumption but revealed neither motives nor subsequent approach. Pairing records with participant accounts could distinguish information requests, discomfort, and withdrawal intentions without classifying every pause as avoidance.

Timing, control, and delivery understanding received more favourable ratings than duration and minimal disruption. Relevant support can still impose excessive time or interaction demands.

Participants' accounts suggested that continuation controls competed with breathing guidance, illustrating this tension in practice. A clearer rhythmic display could reduce competing demands while retaining interruption access. Completing a module alone does not establish engagement with its intended exercise. Evaluation should assess comfortable participation and subsequent skill use, distinguishing content delivery from the intended therapeutic activity~\cite{slovak2024framework}.

Clinical translation requires controls aligned with practitioners' tasks~\cite{brinkman2010therapist}. Pausing an animation cannot stop treatment: patient requests require clinician acknowledgement. A shared interface should distinguish request receipt, reaching a safe stopping point, and readiness to resume, making responsibilities visible. These distinctions preserve clinicians' ability to respond appropriately to the procedural context rather than treating interruption as instantaneous.

\subsection{Distinguishing Companionship from Therapeutic Guidance}
\label{sec:discussion-companionship}

M4 accounts position companionship as an interpersonal regulatory resource through reduced isolation and perceived presence~\cite{zaki2013interpersonal}. Its non-directive accompaniment complements M2's explanations without supplying procedural information.

Unlike Freeman et al.'s coach, which organized repeated cognitive tasks for fear of heights~\cite{freeman2018automated}, M4 offers presence during E5. It neither elicits beliefs nor guides experiments or reviews learning. Adding coping reflection would change participant activity and system responsibility, requiring evaluation as a new component. For example, asking participants to notice their use of an agreed strategy would introduce a distinct therapeutic activity.

Favourable quotations coexist with M4's lowest descriptive acceptability and appropriateness scores (Section~\ref{sec:system-evaluation-results}). Evaluation should examine individual fit, intrusive speech, and support preferences. Familiarity remains a candidate rather than an established benefit; participant choice could accommodate differing needs. Choosing the form or amount of speech could preserve accompaniment without assuming uniform value.

Perceived oversight also requires clarification: the NPC neither monitored safety nor inferred emotion. Dialogue should communicate accompaniment without implying these capabilities. Exposure-oriented evaluation should distinguish coping attributed to participants' actions, acknowledged clinical support, or imagined companion protection, even when reassurance ratings are similar.

\subsection{Evaluating the Activity as well as the Outcome}
\label{sec:discussion-multimodal}

Physiology complements subjective reports without serving as a ground truth for anxiety. At E1, corrected physiological contrasts favoured intervention, whereas the slightly positive subjective contrast did not meet the corrected threshold. VAS-A concerned the preceding event; physiological windows covered 20 seconds before responses, after support and at differing elapsed times. These differences support evaluating experience and arousal separately~\cite{kreibig2010autonomic,posada2020eda}.

Lower HR or EDA cannot establish changed beliefs or transferable skills. Head orientation cannot establish ocular attention, avoidance, or belief testing; region availability and the control condition's absent companion constrain comparisons. Increased realism can intensify threat without clarifying the intended participant activity. E4's symbolic cue does not reproduce pain, and support cannot replace clinical symptom communication.

Evaluation should match the claimed contribution~\cite{slovak2024framework}: assess explanations through comprehension, breathing through participation and later use, and exposure through predictions, subsequent approach, and follow-up. Each measure would address the specific activity that the interaction is intended to support. These measures complement anxiety and acceptability while making the proposed interaction-to-outcome relationship testable.

Future comparisons should separate bundled choices within this fixed, non-adaptive schedule. Physiological data currently neither select support nor determine when it is delivered. Timing tests could vary delivery position while matching content and duration; content tests could compare modules within one event while retaining shared controls. Watkins et al.'s factorial CBT trial provides a methodological precedent, rather than evidence about dental modules~\cite{watkins2023ingredients}. Its component comparisons illustrate how individual contributions can be examined within a broader programme. Repeated and follow-up assessment could then clarify how particular interactions support coping across encounters.

\subsection{Limitations}
\label{sec:limitations}

\paragraph{Design and causal attribution.}
The fixed E1--E5 order confounds event identity with temporal position, accumulated exposure, carryover, habituation, and fatigue. Each event also had one predefined module. Event-specific contrasts therefore concern configurations at particular sequence positions, rather than independent event or module effects, including a separate effect of M1 at E4. Because the runs differed in duration, the event ratings and pre-VAS physiological windows occurred at different elapsed times and under different exposure histories. The observed differences may therefore reflect
support content together with natural recovery, habituation, and accumulated
exposure; they do not isolate the effect of support content from elapsed time. The design evaluates the complete package; it does not establish superiority over continuous distraction or differently timed delivery of the same content. Duration-matched comparisons and counterbalanced event vignettes could address these questions where clinical coherence permits.

\paragraph{Sample and statistical inference.}
The 24-participant sample limits precision, and random allocation produced different age and gender distributions. The subjective model adjusted for baseline current VAS-A and MDAS; physiological contrasts were unadjusted and used outcome-specific complete-case subsets. Bounded, unevenly distributed ratings and approximate model-based tests further constrain inference. Pointwise bootstrap intervals and Holm-adjusted tests answer different questions: an interval excluding zero does not establish a multiplicity-adjusted event difference. Secondary session-level summaries do not replace the primary event-specific comparison.

\paragraph{Measurement validity and missing records.}
Head-orientation availability differs from on-screen visibility and validated tracking time. Geometry, occlusion, sensory titration, and event duration affect its interpretation. AIM and IAM assess acceptability and appropriateness, not relative effectiveness; their translated, module-referenced application was not independently validated. The five TFA-informed diagnostic items are researcher-developed items, rather than a validated composite scale. The perceived-control item included reducing stimulation, although M0 offered pause, resume, and exit without a separate participant-operated intensity control. Agreement therefore does not establish direct control over stimulation intensity. M3 ratings also combine its visual and auditory implementations.

\paragraph{Qualitative coverage.}
Analysis included all 12 intervention transcripts but only seven control transcripts. Short interviews and incomplete control coverage limit thematic breadth and condition comparisons; we do not claim saturation or frequencies of agreement. Retrospective notes from three noisy recordings were excluded and cannot support verbatim quotations. These limits are especially relevant when interpreting selected positive accounts alongside variation in the questionnaire distributions.

\paragraph{Clinical scope and tolerability.}
Eligibility required reported dental fear or a dental visit within six months, rather than an upcoming third-molar extraction. The findings concern immediate experience in simulated dental VR, rather than clinical extraction outcomes or long-term anxiety management. No reported adverse events and nonzero VRSQ symptoms describe different aspects of tolerability. Neither establishes absence of discomfort, a between-group safety advantage, or general system safety.

\section{Conclusion}

We investigated dental-anxiety support organized around specific moments of a simulated procedure. A three-phase formative study informed five Anxiety Events, an intervention-module framework, and four design goals. These informed a standardized event-contingent VR system with fixed event--module assignments and participant-controlled pause and exit functions.

In a randomized study with 24 participants, the intervention condition had lower adjusted event-averaged subjective anxiety than no-intervention VR. The intervention-minus-control difference was $-12.83$ VAS-A points, and E4 and E5 comparisons remained significant after Holm correction. Complementary physiological, behavioural, questionnaire, and interview findings identified opportunities and tensions in how support was experienced.

The design implications emphasize relevant support, preserved agency, manageable interaction demands, and a comprehensible social role for companionship. The findings motivate evaluation of alternative timing and event--module configurations. They concern immediate simulated experiences and establish neither independent module effects nor outcomes during actual dental treatment.

\begin{acks}
This work was supported by the Shenzhen Medical Research Fund (B2503005); NSFC (Grant No.~72495131); the Major Frontier Exploration Program (Grant No.~C10120250085) of the Shenzhen Medical Academy of Research and Translation (SMART); the International Science and Technology Cooperation Center, Ministry of Science and Technology of China (Grant No.~2024YFE0203000); the Guangdong Provincial Key Laboratory of Mathematical Foundations for Artificial Intelligence (2023B1212010001); the Program for Guangdong Introducing Innovative and Entrepreneurial Teams (Grant No.~2023ZT10X044); the Shenzhen Stability Science Program 2023; the Shenzhen Key Lab of Multi-Modal Cognitive Computing; the Shenzhen Science and Technology Program (ZDSYS20230626091302006); and the Shenzhen Loop Area Institute (Grant No.~2026SFP004).
\end{acks}

\section*{The use of AI}
Large language models (GPT-5.6 and GPT-6) were solely used for minor language editing, including grammar and readability improvements.

\clearpage
\appendix
\section{Formative Study Details}
\label{app:formative-details}

This appendix reports methodological and design details moved from Section~\ref{sec:formative} to keep the main paper focused on the formative logic and the evidence required to follow the subsequent system design. It covers participant composition, activity materials, the Phase~1 analytic workflow and full theme inventory, Phase~2 expert design refinements, Phase~3 prototype versions and feedback-to-design relationships, and cross-phase evidence boundaries. The three phases served different evidential roles and should not be interpreted as independent replications of one another.

\subsection{Participants, Recruitment, and Study Roles}
\label{app:formative-participants}

Table~\ref{tab:formative-overview} summarizes phase-specific sample sizes,
activities, and outputs. The recruitment and eligibility details below
supplement that overview.

Phase~1 participants were recruited through a university-affiliated dental
hospital in southern China. Patients were at least 18 years old and had
attended a dental visit within the previous six months; dental anxiety or
fear was not required. Dental students were enrolled in a dentistry programme:
eight were undertaking clinical internships, three were postgraduate students
in residency training, and one was another postgraduate student. The latter
four were also practising dentists with patient-contact experience. The
healthcare-professional group included nine practising dentists, one dental
nurse, one clinical instructor, and one participant working in private
healthcare. The dentists and dental nurse had participated in clinical care;
familiarity with impacted third-molar extraction was not required for this
recruitment group. Recruitment groups therefore did not imply mutually
exclusive professional identities.

The three Phase~2 experts were experienced outpatient dentists recruited through the same
hospital, all of whom had participated in Phase~1. Phase~3 patients were
recruited through advertisements in the local community and on a university
campus and had attended a dental visit within the previous six months. One
had participated in Phase~1; the other 11 had not. Each Phase~3 participant
received RMB~50. Phase-specific counts should not be summed as independent
samples. Phase~1 patient quotation identifiers retain the original participant
number, padded to two digits (e.g., P1-P06).

\subsection{Study Materials and Activity Scope}
\label{app:formative-materials}

Table~\ref{tab:app-formative-materials} summarizes the materials and discussion scope used across the three phases.

\begin{table*}[t]
\caption{Topics and materials used in the formative study.}
\label{tab:app-formative-materials}
\centering
\small
\begin{tabularx}{\textwidth}{@{}p{0.20\textwidth}X@{}}
\toprule
\textbf{Activity} & \textbf{Topics or materials} \\
\midrule
Phase~1 interviews & Anxiety-provoking dental situations and their timing; associated thoughts, sensations, attention, and responses; desired forms of support; and considerations for representing these experiences in VR. One dentist responded to the same guide in writing. \\
Phase~2 participatory design & Anxiety Event cards, Module cards, a system-flow diagram, and an A2 design sheet. The three dentists sorted and combined cards and modified the workflow together. \\
Phase~3 interviews & Baseline accounts of dental experience and pressure points, followed by comments on the experienced events, sequence, comprehensibility, perceived tolerability, and possible improvements after the no-intervention walkthrough. \\
\bottomrule
\end{tabularx}
\end{table*}

\subsection{Phase 1: Interview Procedure, Analysis, and Full Theme Inventory}
\label{app:phase1-analysis}

\subsubsection{Interview Procedure}

Phase~1 interviews were semi-structured and conducted in Chinese. Most synchronous interviews took place online via Tencent Meeting; one dental student joined via Zoom. Because of scheduling constraints, one dentist completed a written questionnaire containing the same questions as the interview guide instead of a synchronous interview. The synchronous interviews were audio-recorded and transcribed verbatim; no concurrent field notes were taken. Quotations in the paper are English translations.

The guide covered four domains. First, participants described situations that elicited dental anxiety and identified when each situation became threatening. Second, they described their thoughts, sensations, attention, and behavioural responses during those situations. Third, they discussed what forms of support might be appropriate at each moment. Finally, they considered how such support could be represented within a simulated dental VR experience.

\subsubsection{Qualitative Analysis}
\label{sec:phase1-analysis}

Two researchers analyzed the Phase~1 materials in Miro using a three-round hybrid deductive--inductive process. The written response was handled as a textual interview record and entered into the same workflow as the synchronous interviews. Before coding, each transcript or written response was segmented into semantic units. After removing non-substantive fillers, retained source excerpts were represented as green \emph{Quote} cards. The researchers then interpreted and condensed each Quote into a concise yellow \emph{Statement} card. Segments with uncertain transcription or interpretation were marked as black cards for joint review.

In Round~1, cards were grouped inductively within each participant to retain individual context. In Round~2, the researchers compared and reorganized material across participants and stakeholder groups. This produced 18 cross-participant themes: seven anxiety-experience themes (H01--H07), eight support-need themes (H08--H15), and three initial system-consideration themes (H16--H18). These counts describe the resulting thematic organization rather than predetermined coding quotas or prevalence estimates. In Round~3, the researchers used the interview questions as a deductive, design-oriented organizing framework: what makes dental situations threatening, what support is sought, and what the system should accommodate. Through joint discussion, they consolidated the themes into T1--T3, S1--S5, concrete anxiety-provoking situations, and cross-cutting system considerations for Phase~2.

The two researchers divided the initial coding workload equally and then cross-checked one another's cards, ensuring that both reviewed the complete Phase~1 dataset. Disagreements were resolved through discussion; unresolved cases would have been referred to a third researcher for adjudication. The CVM was used as an interpretive lens for vulnerability-related appraisals, while Gross's Process Model was used later to organize the regulatory roles of support in the system-design rationale.

\begin{table*}[t]
\caption{Analytic progression from Phase~1 source material to the design inputs carried into Phase~2.}
\label{tab:app-analytic-progression}
\centering
\small
\begin{tabularx}{\textwidth}{@{}p{0.18\textwidth}X p{0.27\textwidth}@{}}
\toprule
\textbf{Stage} & \textbf{Analytic activity} & \textbf{Output or role} \\
\midrule
Data preparation & Segment transcripts and the written response into semantic units; remove non-substantive fillers; distinguish source excerpts from researcher-condensed statements; jointly review uncertain segments. & Quote cards, Statement cards, and cards requiring clarification. \\
Round~1 & Group cards thematically within each participant. & Participant-level organization retaining individual context. \\
Round~2 & Compare and reorganize cards across participants and stakeholder groups. & 18 cross-participant themes: seven anxiety experiences, eight support needs, and three initial system considerations (H01--H18). \\
Round~3 and consolidation & Synthesize themes deductively around the interview questions concerning threats, desired support, and system considerations. & T1--T3, S1--S5, concrete situations, and initial system considerations for Phase~2. \\
Phase~2 translation & Situate the concepts within the selected dental scenario through joint expert design. & Five preliminary Anxiety Events, M0--M4, and a preliminary system workflow. \\
\bottomrule
\end{tabularx}
\end{table*}

Patients' first-person accounts, students' training-related perspectives, and professionals' clinical observations were retained as complementary sources. Their inclusion does not make lived experience interchangeable with clinical interpretation. Threat categories describe \emph{why} a situation is difficult; concrete situations identify \emph{moments} of care; candidate intervention approaches describe \emph{forms of support}; and scenario-specific events and implemented modules are later design translations.

\subsubsection{Illustrative Excerpts}

The main paper includes selected quotations to ground the threat and support categories; additional representative excerpts from the original analysis are retained here.

\begin{itemize}
    \item \textbf{Procedural uncertainty:} ``The most stressful part is waiting without knowing what the dentist will do next. I keep wondering when it will start and how long it will last.'' (P1-P10, patient).
    \item A dental student highlighted the information gap between clinical and patient perspectives: ``We understand what the next action is for, but when patients see the same action, they may not know what will happen next.'' (dental student).
    \item \textbf{Sensory and invasive threat:} ``When I see the needle coming toward me, I tense up before it even touches me. The drill sound and vibration make the whole procedure feel more threatening.'' (P1-P09, patient).
    \item \textbf{Expectation violation:} ``After the anaesthetic, or when the drill stops, I expect the difficult part to be over. If I then feel pain or they start again, the anxiety comes back immediately.'' (P1-P04, patient).
    \item A patient described completion expectation: ``I thought that when the drill stopped, it was over, but I would feel more nervous when the dentist picked it up again.'' (patient). A dentist contextualized this as clinically plausible variability: ``Sometimes dental surgery does not follow a completely fixed sequence, and it can be quite normal for additional work to be needed.'' (dentist).
    \item \textbf{Agency and guidance:} ``I would feel safer if I could raise my hand and know the dentist would pause. It would also help to hear what the next step is before it happens.'' (P1-P06, patient).
    \item \textbf{Momentary regulation:} ``When the drill starts, a short breathing prompt or turning the sound down a little would help me calm down without taking me out of the procedure.'' (P1-P06, patient).
    \item \textbf{Interpersonal support:} ``Even a simple message that I was doing well and could pause if needed would make me feel that someone was paying attention to how I felt.'' (P1-P02, patient).
\end{itemize}

\subsubsection{Theme Inventory and Consolidation Rules}

Tables~\ref{tab:app-themes-anxiety}--\ref{tab:app-themes-system} report the 18 Phase~1 themes. The scope descriptions summarize thematic content; they are not additional participant quotations or frequency estimates. H06 requires an established expectation of numbness followed by an inconsistent sensation; anticipating pain before treatment without this temporal mismatch is not equivalent to expectation violation. Likewise, H07 concerns renewed operation after expected completion rather than duration alone. H05 captures vibration and pressure in accounts of dental care, whereas the implemented E3 realizes drilling onset through audio and does not reproduce every sensory modality represented within T2.

\begin{table*}[t]
\caption{Phase~1 anxiety-experience themes (H01--H07).}
\label{tab:app-themes-anxiety}
\centering
\small
\begin{tabularx}{\textwidth}{@{}p{0.06\textwidth}p{0.27\textwidth}X p{0.08\textwidth}@{}}
\toprule
\textbf{ID} & \textbf{Theme} & \textbf{Scope} & \textbf{Role} \\
\midrule
H01 & Waiting and Temporal Uncertainty & Not knowing when treatment will begin, waiting during preparation, or being unable to judge how long the current step will last. & T1 \\
H02 & Uncertainty about Procedural Actions and Their Meaning & Not knowing the next action, being unable to understand an observed action's purpose, or being unable to identify the current procedural stage. & T1 \\
H03 & Visibility and Proximity of Invasive Instruments & Seeing needles, drills, or other invasive instruments approach the face and feeling threatened before contact occurs. & T2 \\
H04 & Threat Associated with Drilling Sounds & The sudden onset or salience of drilling sounds, or using sound to infer that an operation is about to begin or is continuing. & T2 \\
H05 & Vibration, Pressure, and Perceived Bodily Intrusion & Vibration or pushing sensations, a sense of instruments acting within the body, and uncertainty about what these sensations mean. & T2 \\
H06 & Unexpected Sensations after Expected Numbness & Pain or discomfort that conflicts with an established expectation of numbness, prompting uncertainty about anaesthesia or the ongoing procedure without establishing a clinical cause. & T3 \\
H07 & Violation of Expected Procedural Completion & Expecting completion when drilling stops or an instrument moves away, followed by learning that further operation is required. & T3 \\
\bottomrule
\end{tabularx}
\end{table*}

\begin{table*}[t]
\caption{Phase~1 support-need themes (H08--H15).}
\label{tab:app-themes-support}
\centering
\small
\begin{tabularx}{\textwidth}{@{}p{0.06\textwidth}p{0.27\textwidth}X p{0.08\textwidth}@{}}
\toprule
\textbf{ID} & \textbf{Theme} & \textbf{Scope} & \textbf{Role} \\
\midrule
H08 & Nonverbal Communication of Discomfort & Expressing discomfort without speaking, for example by raising a hand, and having that signal recognized and acknowledged. & S1 \\
H09 & Control over Pausing, Stopping, and Pacing & Being able to pause or end the experience without waiting for a step to finish, and understanding when the experience will resume. & S1 \\
H10 & Breathing Guidance and Brief Self-regulation & Brief breathing cues, a slower breathing rhythm, or a short opportunity to regain composure before continuing. & S2 \\
H11 & Timely Explanations of Current and Upcoming Steps & Explanations near the relevant action of what is happening, what comes next, and why, rather than information provided only at the beginning. & S3 \\
H12 & Preparation for Expected Sensations and Progress & Anticipating sounds and sensations, understanding transitions, and distinguishing a temporary pause from procedural completion. & S3 \\
H13 & Attenuation of Threatening Visual Cues & Reducing instrument detail through local masking or decreased visual prominence of needles and drills. & S4 \\
H14 & Modulation of Threatening Auditory Cues & Reducing the intensity or salience of drilling sounds while retaining necessary procedural or communication cues. & S4 \\
H15 & Reassurance, Recognition, and Supportive Presence & Having anxiety acknowledged, feeling that someone attends to one's experience, and receiving calm, non-directive reassurance or companionship. & S5 \\
\bottomrule
\end{tabularx}
\end{table*}

\begin{table*}[t]
\caption{Phase~1 cross-cutting system considerations (H16--H18). These themes constrain the complete experience and do not constitute an additional intervention approach.}
\label{tab:app-themes-system}
\centering
\small
\begin{tabularx}{\textwidth}{@{}p{0.06\textwidth}p{0.28\textwidth}X p{0.24\textwidth}@{}}
\toprule
\textbf{ID} & \textbf{Theme} & \textbf{Scope} & \textbf{Design role} \\
\midrule
H16 & Contextual Comprehensibility and Procedural Coherence & Understanding the scene, matching instrument actions with sounds, and avoiding waits or transitions that resemble a stalled application. & Constraints on scenario, event order, and transitions. \\
H17 & Low-burden and Minimally Disruptive Support & Limiting text and competing prompts during anxiety without fully obscuring the ongoing procedure. & Constraints on information length, position, combinations, and timing. \\
H18 & Understandable Controls and Predictable System Responses & Knowing how to operate the system, what follows a pause request, and why support appears and when it ends. & Input to instructions, state feedback, and transparency. \\
\bottomrule
\end{tabularx}
\end{table*}

\paragraph{From threats to events.}
H01--H02 inform T1 and its later realization as uncertainty about an imminent procedure in E1. H03 informs the visual instrument exposure of E2. H04--H05 retain the distinction between auditory and bodily threats within T2, while the implemented E3 specifically realizes drilling onset through audio. H06 and H07 inform the expectation mismatches represented in E4 and E5. These relationships connect Phase~1 threat analysis with later scenario translation; they do not imply that E1--E5 were finalized during Phase~1 or that each theme required a separate event.

\paragraph{From support needs to modules.}
H08--H09 inform S1 and the M0 foundation; H10 informs S2 and M1; H11--H12 inform S3 and M2; H13--H14 inform S4 and M3; and H15 informs S5 and M4. Within S1, signalling discomfort is distinct from the control and acknowledgement that should follow. Brief breathing regulation under S2 is likewise distinct from an independently available right to pause. The visual and auditory needs within S4 need not be addressed simultaneously, and interpersonal reassurance belongs to S5 even when delivered through the same spoken channel as procedural guidance. These functional distinctions informed module construction, not the later fixed event--module assignments.

H16--H18 constrain comprehensibility, support burden, and predictability across the experience. The threat--event and approach--module relationships are separate design translations; neither determines the final event--module pairing, which the research team specified during system design. A source account can inform more than one conceptual relationship when it expresses distinct needs, so these mappings should not be read as exclusive assignments of source excerpts.

\subsection{Phase 2: Participatory-Design Procedure and Expert Refinements}
\label{app:phase2-details}

The three returning outpatient dentists jointly participated in one approximately 30-minute participatory-design session. They directly sorted and combined cards and modified the system workflow on a shared A2 design sheet. The card set and flow diagram were derived from Phase~1 and represented anxiety-provoking moments and candidate support approaches organized around T1--T3 and S1--S5. The experts were asked to identify a bounded dental procedure in which the materials could be represented coherently, place clinically plausible anxiety-provoking moments within that procedure, review the appropriateness and timing of candidate strategies, and organize the elements into a preliminary system workflow.

The activity evaluated clinical plausibility and functional appropriateness rather than therapeutic efficacy. The dentists considered whether each proposed moment was plausible within the selected procedure, whether a corresponding support addressed the dominant threat, whether its timing or intensity could unnecessarily disrupt the procedure, and whether the workflow preserved patient agency and safety. Written meeting notes documented the activity alongside the shared design artifact. The cards and system-flow diagram were materials for joint construction rather than only items for approval.

The dentists recommended unilateral extraction of an impacted mandibular third molar under local anaesthesia. Restricting the simulation to one mandibular side also supported a stable patient viewpoint, reproducible instrument trajectories, and consistent event/intervention timing. Crossing the three Anxiety Threats with the staged scenario produced five preliminary events: E1 (unknown next step before treatment), E2 (visual appearance/proximity of invasive instruments), E3 (abrupt drilling-sound onset), E4 (unexpected pain-related cue after expected numbness), and E5 (renewed treatment after cues suggesting apparent completion).

Table~\ref{tab:app-expert-refinements} records six design issues and the associated expert refinements. In EX5, the expert input concerned competition between supportive content; the exact fixed event--module mapping remained a subsequent research-team decision.

\begin{table*}[t]
\caption{Expert input and resulting Phase~2 refinements. Initial issues refer to candidate proposals, not necessarily features implemented in the patient-facing prototype.}
\label{tab:app-expert-refinements}
\centering
\small
\begin{tabularx}{\textwidth}{@{}p{0.06\textwidth}p{0.34\textwidth}X@{}}
\toprule
\textbf{ID} & \textbf{Initial issue} & \textbf{Expert input and resulting refinement} \\
\midrule
EX1 & Pausing, breathing, and explanations were grouped as support offered after an event. & Discomfort could occur outside predefined support moments. Pause and stop controls were therefore organized as an always-available M0 foundation, distinct from M1--M4. \\
EX2 & Comprehensive information about instruments, steps, and sensations was placed before the simulation, with no further explanations during it. & Information should be linked to the action currently unfolding. Preparatory information was separated from brief explanations of current and upcoming steps, informing M2. \\
EX3 & Candidate sensory-reduction approaches obscured the entire view or removed all sound. & Reducing threatening stimulation should not remove necessary context. M3 was refined toward selective attenuation of instrument detail or drilling sounds while retaining essential scene and communication cues. \\
EX4 & The candidate event attributed unexpected sensations after anticipated numbness to anaesthetic failure. & The experts cautioned against assigning a clinical cause to the patient's experience. E4 was reframed as an unexpected pain-related cue after expected numbness, preserving expectation violation without diagnosing its cause. \\
EX5 & Breathing prompts, procedural explanations, visual changes, and supportive speech were proposed concurrently. & Competing support content should be reduced and support relevant to the immediate difficulty prioritized. The research team subsequently determined the final one-module-per-event mapping; M0 remained independently available. \\
EX6 & Candidate reassurance promised no pain or imminent completion, or instructed patients not to feel anxious. & Support should acknowledge feelings without guaranteeing outcomes or treating anxiety as inappropriate. M4 was refined toward calm, non-directive reassurance and supportive presence; procedural explanations remained within M2. \\
\bottomrule
\end{tabularx}
\end{table*}

The resulting Phase~2 output comprised three linked design elements: a bounded clinical scenario, five preliminary event representations, and an M0--M4 module framework within a preliminary workflow. The clinical scenario supplied procedural context; the event sequence specified moments to represent; and the module framework specified the forms of support to implement. The precise event--module assignments and delivery rules were subsequently determined by the research team in Section~\ref{sec:system}.

\subsubsection{Expert Participatory-Design Artifact}
\label{app:phase2-artifact}

\begin{figure*}[htbp]
  \centering
  \IfFileExists{Figures/Phase2_A2_design.pdf}{
    \includegraphics[width=\textwidth,height=0.65\textheight,keepaspectratio]{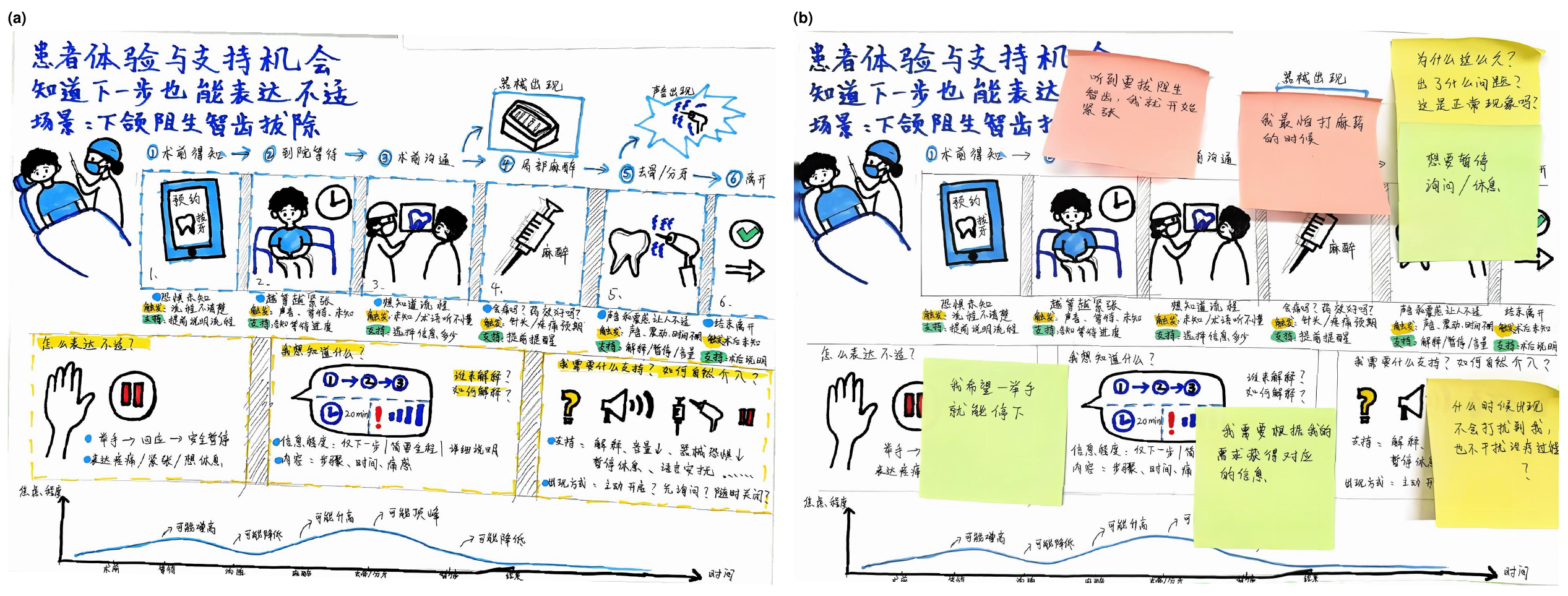}
  }{
    \fbox{\parbox[c][0.23\textheight][c]{0.94\textwidth}{\centering Expert participatory-design artifact\\[0.6em]Shared A2 design sheet}}
  }
  \caption{Shared A2 design artifact from Phase~2. Three outpatient dentists jointly sorted and combined the design cards and modified the system workflow.}
  \Description{A scanned shared A2 design sheet documenting the card-based participatory-design activity with three dentists.}
  \label{Figure:Phase2_A2_design}
\end{figure*}

The shared A2 sheet captures the output of the joint card-sorting, combination, and workflow-design activity. Together with the meeting notes, it documents the design work undertaken by the three dentists (Figure~\ref{Figure:Phase2_A2_design}). The artifact complements the scenario, event, and module specifications reported in Section~\ref{sec:formative-phase2}.

\subsection{Phase 3: Walkthrough Protocol and Patient-Informed Refinements}
\label{app:phase3-walkthrough}

All 12 Phase~3 patients experienced the same preliminary no-intervention
prototype. Feedback from the full cohort was collected before any revisions
were implemented; subsequent changes were not retested by this cohort.
Recruitment and participant overlap are reported in
Appendix~\ref{app:formative-participants}.

Each session began with a semi-structured interview about the participant's
dental-anxiety experiences and pressure points. Participants then used a Meta
Quest~3 for an approximately 10-minute walkthrough containing the complete
preliminary E1--E5 sequence. They could raise their left hand to request a
pause, and a researcher remained beside them throughout to respond. The
prototype and its differences from the later system are documented in
Table~\ref{tab:phase3-refinements}.

After the walkthrough, participants discussed each event's plausibility as
an anxiety-provoking dental moment, the comprehensibility of the sequence
and transitions, perceived tolerability and safety, and needed scene or
workflow changes. The interview component lasted approximately 15 minutes
per participant. The interview and post-walkthrough material underwent the
card preparation, three-round hybrid deductive--inductive analysis, and
cross-checking procedure described in
Appendix~\ref{sec:phase1-analysis}. Round~1 retained each participant's
experienced event sequence; Round~2 compared feedback across participants;
and Round~3 organized it around event face plausibility, coherence and
comprehensibility, perceived tolerability and safety, and workflow
refinements. Phase~2 meeting notes and revised artifacts remained separate
design-decision evidence and were not included in this interview coding.

The research team retained the fixed E1--E5 sequence and refined the
transitions between events as summarized in
Table~\ref{tab:phase3-refinements}. The entries distinguish intended dental
experiences from artifacts of their VR representation; they are analytic
summaries, not verbatim quotations or counts of agreement. No participant
discontinued the supervised walkthrough because of discomfort. This records
completion under the procedure described above, rather than absence of
discomfort or general safety of the final system.

\begin{table*}[htbp]
  \caption{The common Phase~3 prototype, patient interpretations, and
  subsequent refinements. All 12 patients experienced the same
  no-intervention prototype. The last column records later design decisions,
  not a second validation.}
  \label{tab:phase3-refinements}
  \small
  \setlength{\tabcolsep}{5pt}
  \renewcommand{\arraystretch}{1.13}
  \begin{tabularx}{\textwidth}{@{}>{\raggedright\arraybackslash}p{0.12\textwidth}>{\raggedright\arraybackslash}X>{\raggedright\arraybackslash}X@{}}
    \toprule
    Component & Experienced prototype and patient interpretation
      & Subsequent design response and scope \\
    \midrule
    E1: Waiting
      & Treatment was understood to be imminent while its onset remained
        uncertain. Inactivity without preparation cues could resemble a
        stalled application.
      & Retain visible preparation and a bounded wait with limited next-step
        information, preserving basic scene comprehension. \\
    \addlinespace
    E2: Instruments
      & Instrument approach could be threatening before contact or drilling;
        abrupt, excessively close, or implausible motion could foreground
        VR presentation artifacts.
      & Refine approach trajectory, distance, and visible duration; retain
        a distinct visual-exposure stage before drilling begins. \\
    \addlinespace
    E3: Drilling audio
      & Drilling audio marked the transition to operation. Simultaneous cue
        onset made instrument approach and operation harder to distinguish.
        E3 used no vibration or pressure stimulus.
      & Establish instrument visibility before abrupt drilling onset and
        synchronize sound with the operation animation. E3 remained
        audio-based; auditory sensory titration was added later as M3.2. \\
    \addlinespace
    E4: Unexpected pain-related cue
      & Virtual-dentist actions and narration established local anaesthesia
        and an expectation of numbness before Meta Quest~3 controller
        vibration. Patients understood the vibration as a bodily cue
        symbolizing pain; its unexpectedness depended on that context.
        The prototype had no peripheral red gradient.
      & Retain the anaesthesia context and symbolic controller cue without
        attributing a clinical cause. Add a peripheral red gradient after
        cohort feedback to reinforce the intended pain-related meaning.
        This representation does not reproduce oral pain. \\
    \addlinespace
    E5: Continued treatment
      & The dentist explicitly conveyed completion before further operation.
        Renewed treatment conflicted with patients' expectations of the
        dentist's actions and increased tension.
      & Remove the explicit completion statement. Use drilling cessation,
        instrument withdrawal, and a brief pause to suggest completion;
        then explain the additional operation and resume drilling. \\
    \addlinespace
    Pause mechanism
      & A left-hand signal requested a researcher-mediated pause. The final
        in-VR pause menu was absent.
      & Add the interaction tutorial and application-controlled M0 pause,
        resume, and exit functions. \\
    \addlinespace
    M1--M4
      & Phase~2 proposed the module framework, but no modules were delivered
        during the walkthrough.
      & The research team subsequently implemented the fixed event--module
        assignments (Section~\ref{sec:event-module-sync}). Evaluation of
        the intervention package belongs to the controlled user study. \\
    \bottomrule
  \end{tabularx}
\end{table*}

\subsection{Cross-Phase Evidence Boundaries and Traceability}
\label{app:evidence-boundaries}

Table~\ref{tab:app-evidence-boundaries} summarizes what each formative source can and cannot support. This distinction is especially important because the final event--module assignments and some post-walkthrough revisions were research-team synthesis decisions informed by, but not directly validated within, the formative phases.

\begin{table*}[t]
\caption{Relationship between formative-study materials, design outputs, and evidence boundaries.}
\label{tab:app-evidence-boundaries}
\centering
\scriptsize
\begin{tabularx}{\textwidth}{@{}p{0.20\textwidth}p{0.25\textwidth}X@{}}
\toprule
\textbf{Claim or design output} & \textbf{Relevant source} & \textbf{Scope of interpretation} \\
\midrule
T1--T3 and S1--S5 & Phase~1 transcripts, written response, and Miro cards. & Characterizes recurring threats and support needs, not clinical diagnoses, prevalence, or agreement among stakeholder groups. Themes, concrete situations, and subsequent VR events are different analytic/design levels. \\
Scenario, module framework, and preliminary workflow & Phase~2 meeting notes and jointly revised cards/A2 workflow; three returning Phase~1 dentists. & Supports clinical plausibility and functional design decisions. The activity is not an independent expert replication, a test of therapeutic efficacy, or the source of all subsequent implementation choices. \\
Event interpretations and patient-facing refinements & Phase~3 interviews after all 12 participants experienced the same no-intervention prototype. & Supports interpretation of the experienced scene and identifies revision needs. It does not establish clinical equivalence, anxiety-reduction effects, or validation of M1--M4, final in-VR controls, or the complete intervention workflow. Controller vibration was interpreted as a bodily cue representing pain, not dental-treatment pain. \\
Supervised walkthrough tolerability & No participant discontinued because of discomfort; a researcher remained beside participants to respond to the left-hand pause signal. & Describes completion under this supervised procedure, not absence of discomfort or general safety of the system and later-added features. \\
Refined events, DG1--DG4, and event--module assignments & Research-team synthesis of formative inputs and the theoretical design rationale; revisions followed cohort feedback. & Specifies the design selected for implementation. It does not imply that later changes, including the E4 red gradient and revised E5 transition, were retested by the Phase~3 cohort. Event distinctions and fixed mappings do not establish independent causal effects of events or modules. \\
\bottomrule
\end{tabularx}
\end{table*}

The four design goals reported in Section~\ref{sec:formative-goals} therefore represent a cross-phase synthesis rather than the output of a single phase. H16--H18 supplied constraints on coherence, support burden, and predictable interaction; expert refinements clarified agency, contextual information, selective sensory attenuation, and non-directive reassurance; and patient feedback informed event presentation and transitions. Their implementation and evaluation are described in the subsequent system and user-study sections.

\section{User Study: Supplementary Quantitative Reporting}
\label{app:study-quantitative}

The tables in this appendix report the analyses presented in Section~\ref{sec:results}. Counts refer to the records available for each analysis. All group differences are intervention minus control. No demographic balance tests, module-effect tests, or additional tests of session-level questionnaire differences were performed for these summaries.

\begin{table}[htbp]
\caption{Participant characteristics and session-level questionnaire
  summaries. Categorical variables are reported as counts and continuous
  scores as mean (SD). Age was recorded in bands, so an exact mean age is
  unavailable.}
\label{tab:study-descriptives}
\centering
\small
  \setlength{\tabcolsep}{6pt}
  \renewcommand{\arraystretch}{1.02}
  \begin{tabularx}{\textwidth}{@{}>{\raggedright\arraybackslash}Xrr@{}}
    \toprule
    Characteristic or measure & Intervention ($n=12$) & Control ($n=12$) \\
    \midrule
    \multicolumn{3}{@{}l}{\textit{Age, years}} \\
    \quad 18--24 & 5 & 11 \\
    \quad 25--34 & 3 & 0 \\
    \quad 35--44 & 3 & 0 \\
    \quad 45--54 & 1 & 1 \\
    \addlinespace[3pt]
    \multicolumn{3}{@{}l}{\textit{Gender}} \\
    \quad Men & 7 & 4 \\
    \quad Women & 5 & 8 \\
    \addlinespace[3pt]
    \multicolumn{3}{@{}l}{\textit{Highest educational attainment}} \\
    \quad Bachelor's degree & 6 & 11 \\
    \quad Master's degree or above & 6 & 1 \\
    \addlinespace[3pt]
    \multicolumn{3}{@{}l}{\textit{Prior VR use}} \\
    \quad Never & 3 & 3 \\
    \quad Once or twice & 9 & 8 \\
    \quad Occasionally (1--3 times per month) & 0 & 1 \\
    \addlinespace[3pt]
    \multicolumn{3}{@{}l}{\textit{Time since most recent dental treatment}} \\
    \quad Within one month & 0 & 1 \\
    \quad 1--3 months & 5 & 2 \\
    \quad 3--6 months & 3 & 3 \\
    \quad 6--12 months & 1 & 1 \\
    \quad One year or more & 2 & 5 \\
    \quad Never received dental treatment & 1 & 0 \\
    \addlinespace[3pt]
    \multicolumn{3}{@{}l}{\textit{Other dental-history items}} \\
    \quad Previous dental examinations and treatment: yes & 11 & 12 \\
    \quad Prior impacted-third-molar extraction: yes & 6 & 4 \\
    \addlinespace[3pt]
    \multicolumn{3}{@{}l}{\textit{Pre-session scores}} \\
    \quad Baseline current VAS-A (0--100) & 10.75 (11.75) & 14.24 (17.34) \\
    \quad MDAS (5--25) & 15.58 (3.29) & 13.25 (4.37) \\
    \addlinespace[3pt]
    \multicolumn{3}{@{}l}{\textit{Post-session scores}} \\
    \quad Current VAS-A (0--100) & 10.17 (10.45) & 18.08 (8.75) \\
    \quad VRSQ total (0--100) & 5.56 (4.58) & 11.46 (12.39) \\
    \quad VRSQ oculomotor (0--100) & 8.33 (6.15) & 14.58 (13.82) \\
    \quad VRSQ disorientation (0--100) & 2.78 (6.00) & 8.33 (12.75) \\
    \bottomrule
  \end{tabularx}
  \par\smallskip
  \begin{minipage}{\textwidth}
    \footnotesize
    Categories with zero records in both groups are omitted. Dental-history
    and treatment-recency rows summarize separate questionnaire items;
    recency category labels retain the administered form's wording.
    Prior impacted-third-molar extraction was not an exclusion criterion.
    Post-session scores are reported separately from pre-session characteristics.
  \end{minipage}
\end{table}

\begin{table}[htbp]
\caption{VAS-A contrasts in points on the 0--100 scale. Every event uses 12 intervention and 12 control records. Confidence intervals are 2,000-resample participant-cluster bootstrap percentile intervals. Auxiliary $p$-values are model-based approximations; Holm correction covers the five event-specific tests. The overall contrast equally weights E1--E5 and is not the reference-event Group coefficient.}
\label{tab:vas-contrasts}
\centering
\small
\begin{tabular}{@{}lrrrr@{}}
\toprule
Contrast & Difference & 95\% CI & $p$ & $p_{\mathrm{Holm}}$ \\
\midrule
E1 & 1.37 & $[-14.16,21.43]$ & .8656 & 1.0000 \\
E2 & 3.38 & $[-15.53,24.88]$ & .6769 & 1.0000 \\
E3 & $-11.89$ & $[-24.66,3.42]$ & .1520 & .4560 \\
E4 & $-32.66$ & $[-43.31,-22.12]$ & .0006 & .0029 \\
E5 & $-24.36$ & $[-36.14,-13.15]$ & .0063 & .0253 \\
\addlinespace
Overall & $-12.83$ & $[-24.42,-1.70]$ & .0399 & --- \\
\bottomrule
\end{tabular}
\end{table}

\begin{table*}[htbp]
\caption{HR is the difference from resting mean HR in bpm; EDA is the natural-log ratio of window to resting mean total EDA. $n$ gives intervention/control records. Intervals use 20,000 participant-cluster bootstrap resamples; $p$-values use 50,000 participant-level group-label permutations. Holm correction is separate for the five events within each outcome. Overall analyses require valid observations at all five events and compare participant-level equal-weight means. Bootstrap intervals are neither simultaneous intervals nor inversions of the permutation tests.}
\label{tab:physiology-contrasts}
\centering
\small
\begin{tabular}{@{}llrrrrr@{}}
\toprule
Outcome & Contrast & $n$ (I/C) & Difference & 95\% CI & $p$ & $p_{\mathrm{Holm}}$ \\
\midrule
HR & E1 & 10/12 & $-4.399$ & $[-5.680,-3.140]$ & .00002 & .00010 \\
   & E2 & 11/12 & $-1.298$ & $[-2.591,-0.011]$ & .07602 & .13668 \\
   & E3 & 11/12 & $-1.250$ & $[-2.460,-0.085]$ & .06834 & .13668 \\
   & E4 & 12/11 & $-4.658$ & $[-6.485,-2.840]$ & .00010 & .00040 \\
   & E5 & 12/12 & $-3.035$ & $[-4.540,-1.419]$ & .00180 & .00540 \\
   & Overall & 8/11 & $-3.009$ & $[-4.226,-1.746]$ & .00016 & --- \\
\addlinespace
EDA & E1 & 11/12 & $-0.0757$ & $[-0.1123,-0.0400]$ & .00084 & .00420 \\
    & E2 & 12/12 & $-0.0377$ & $[-0.0751,-0.0014]$ & .06464 & .12928 \\
    & E3 & 12/12 & 0.0053 & $[-0.0292,0.0381]$ & .77194 & .77194 \\
    & E4 & 12/12 & $-0.0842$ & $[-0.1316,-0.0372]$ & .00254 & .01016 \\
    & E5 & 12/12 & $-0.0780$ & $[-0.1297,-0.0294]$ & .00518 & .01554 \\
    & Overall & 11/12 & $-0.0553$ & $[-0.0907,-0.0214]$ & .00524 & --- \\
\bottomrule
\end{tabular}
\end{table*}

\begin{table}[htbp]
\caption{Each entry summarizes 12 participant-level means, each computed from four 1--5 items. M3 combines its visual and auditory implementations. Differences among modules are descriptive and do not identify relative effectiveness.}
\label{tab:module-descriptives}
\centering
\small
\begin{tabular}{@{}lrrrr@{}}
\toprule
Module & AIM mean & AIM median & IAM mean & IAM median \\
\midrule
M0 & 3.73 & 3.50 & 4.08 & 4.13 \\
M1 & 4.00 & 3.63 & 3.83 & 3.88 \\
M2 & 4.02 & 4.13 & 4.19 & 4.13 \\
M3 & 3.83 & 4.00 & 3.77 & 3.75 \\
M4 & 3.46 & 3.38 & 3.58 & 3.50 \\
\bottomrule
\end{tabular}
\end{table}

\paragraph{Physiological exclusions and scoring details.}
Invalid HR windows were P002--E2, P004--E3, P008--E1, P011--E1, and P018--E4; the invalid EDA window was P008--E1. Figure~\ref{fig:physiological-profiles} marks these cells directly. VRSQ oculomotor and disorientation subscales were calculated as their item sums divided by 12 and 15, respectively, multiplied by 100; the total is the mean of these two percentages. For the five system diagnostic items, the counts at scores 1, 2, 3, 4, and 5 were, respectively: timing $(0,1,1,6,4)$; duration $(0,2,3,6,1)$; minimal disruption $(0,1,3,5,3)$; perceived control $(0,1,1,5,5)$; and understanding delivery $(0,0,2,6,4)$. The minimal-disruption item was positively worded and was not reverse-scored. No total score was calculated across these five items.

\subsection{Statistical Implementation and Sensitivity Analysis}
\label{app:study-analysis-details}

The VAS-A analyses were conducted in R~4.6.1 with \texttt{nlme}~3.1.169. Models were fitted by restricted maximum likelihood. For each linear contrast, the script used the smallest coefficient-level \texttt{nlme} denominator degrees of freedom among coefficients with nonzero contrast weights. This approximation yielded 20 degrees of freedom for adjusted contrasts and 22 for unadjusted contrasts. It was not a Satterthwaite or Kenward--Roger correction. The Group~$\times$~Event result came from a marginal Wald $F$ test with 4 and 88 degrees of freedom.

The adjusted and unadjusted models each returned estimates for all 2,000 participant-level bootstrap resamples, with no failed resamples. Three near-zero random-intercept fits were retained in the adjusted-model bootstrap; the original fitted model was not a boundary fit. Resampled copies of the same participant received distinct cluster identifiers. The effective random seed was 933554218. The 120 ratings included 17 zeros and two values of 100. These bounded, unevenly distributed responses and the small participant sample limit precision and the reliability of approximate inference.

The sensitivity model omitted baseline current VAS-A and MDAS while retaining Group~$\times$~Event and the participant random intercept. Its equally weighted overall difference was $-13.21$ points (95\% bootstrap CI $[-23.74,-2.34]$; auxiliary $p=.0296$). E4 and E5 also met the .05 threshold after Holm correction ($p_{\mathrm{Holm}}=.0019$ and .0188, respectively). Thus, the adjusted and unadjusted analyses produced similar overall estimates and identified the same two event-specific comparisons. We report original-scale effects rather than constructing a standardized effect with an unspecified denominator.

For physiological coverage checks, the denominator was the expected number of samples in each window, not the number of received rows. Coverage was calculated using the verified sampling rate of 128 samples/s, corresponding to 2,560 expected samples over 20~s and 7,680 expected samples over 60~s. Permutation $p$-values used the absolute Welch statistic and $(b+1)/(50{,}000+1)$, where $b$ is the number of permuted statistics at least as large as observed. Bootstrap intervals were computed separately and are not inversions of these permutation tests. Overall physiological contrasts required five valid windows per participant and therefore describe an outcome-specific complete-case subset.

\section{Supplementary Head-Directed Orientation and Pause Summaries}
\label{app:head-orientation}

This supplementary analysis concerns the analyzed Head Gaze and pause records. It provides a descriptive view of orientation across shared scene regions, not eye-tracking or anxiety measurements. The E1--E5 summaries contain 24 participant records, with 12 in each condition. The NPC is not a shared target because it is absent from control. Its availability begins at M4 onset within E5 and ends at E5 offset. This interval begins 15.67--18.30~s after E5 onset across the 12 intervention records. The records do not separately timestamp resumed drilling, so this is not a measured drilling-to-entry delay. Recomputed candidate hits, primary targets, and qualified episodes reflect these availability boundaries; E5 is included in the analysis.

\paragraph{Operational definitions and temporal aggregation.}
The analysis records represent head pose at 15~Hz and specify 31 rays within a 20-degree full cone, with a maximum distance of 5~m. Each ray retained only the nearest collider intersection, excluding fully occluded targets from its candidate hits. Candidate targets were ranked by angular offset from the central head direction, then by hit distance; only one was selected as primary per interval. A qualified episode required four consecutive primary-hit intervals, approximately 0.267~s. Up to two deviating intervals, approximately 0.133~s, could be bridged thereafter. We counted actual primary-hit time within qualified episodes, excluding the bridged intervals themselves. Previously qualified episodes were intersected with exact event boundaries rather than requalified separately within each cropped interval. Confirmation timestamps were kept separate from effective episode boundaries and were not subtracted to estimate qualified duration.

\begin{figure*}[htbp]
  \centering
  \includegraphics[width=\textwidth]{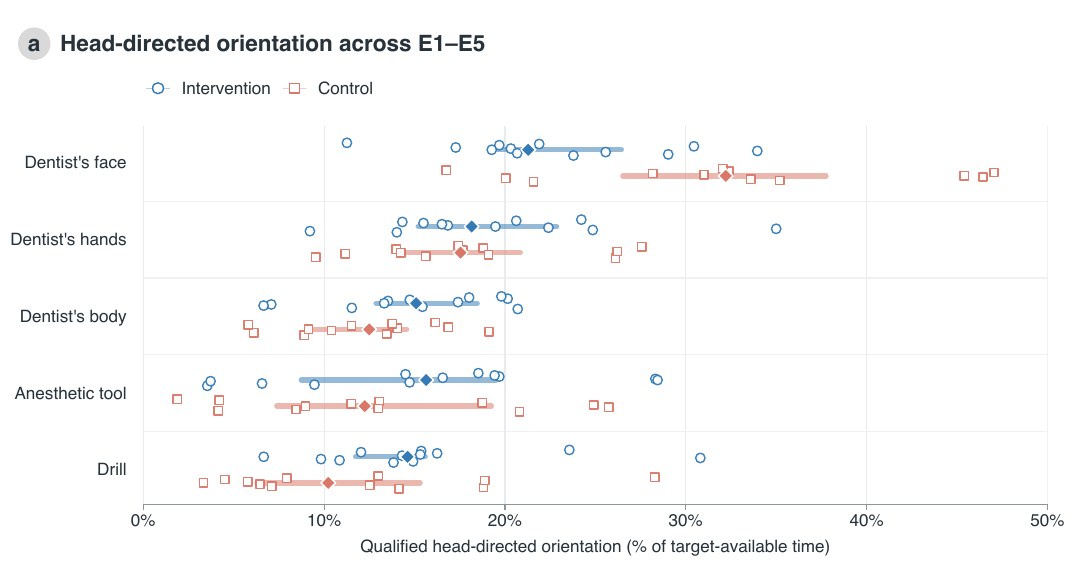}
  \caption{Head-directed orientation across five shared target regions during E1--E5. Each small mark represents one participant record ($n=12$ per condition per target): blue circles denote intervention and coral squares denote control. Diamonds and horizontal bars show medians and interquartile ranges, not confidence intervals. Percentages divide qualified primary-hit time by target-available time, summed within each participant before aggregation. The M2 support period follows the logged E1 offset and is excluded. Module periods within E2--E5 and pauses within the logged event intervals remain included. E5 uses primary-target assignments recomputed with NPC availability restricted to M4 onset through E5 offset. The NPC is described separately because it is absent from control. The plot describes head orientation, not ocular gaze or anxiety, and no hypothesis tests are shown.}
  \Description{A horizontal dot plot compares intervention and control records for the dentist's face, hands, and body, an anaesthetic tool, and a drill across E1 to E5. A gray circle identifies panel a. Blue open circles and coral open squares show individual percentages; filled diamonds mark medians, and horizontal lines span the first to third quartiles. The horizontal axis runs from zero to 50 percent of target-available time. Intervention and control medians are 21.29 and 32.21 percent for the face, 18.16 and 17.54 for the hands, 15.09 and 12.50 for the body, 15.64 and 12.25 for the anaesthetic tool, and 14.62 and 10.23 for the drill. Individual distributions overlap, and no significance markers are present.}
  \label{fig:head-orientation-supplement}
\end{figure*}

For each participant and target, we summed qualified primary-hit seconds within E1--E5 and divided by that target's summed declared available seconds within those events. The dentist's face, hands, and body were available across E1--E5; the anaesthetic tool was available within E2 and E4, and the drill within E2, E3, and E5. The resulting percentages were summarized across participants using medians and interquartile ranges (Figure~\ref{fig:head-orientation-supplement}). Target-unavailable intervals were not converted into zero-orientation observations. The separate M2 interval following E1 was excluded because it lay outside the E1 boundary; module periods nested within E2--E5 and pauses within event boundaries remained included. These whole-event summaries differ from the pre-VAS physiological windows and weight intervals by target-available duration, rather than assigning equal weight to each event.

Declared availability is not equivalent to on-screen visibility, perceptual salience, or validated tracking time. The recorded data contain no independent tracking-validity indicator; a finite pose or a null target does not establish successful tracking or tracking failure. Target geometry, occlusion, sensory titration, and unequal event durations remain relevant to interpretation. Percentages across targets need not sum to 100 because their availability denominators differ. These summaries do not establish independent module effects.

For the NPC, the denominator included only its available interval from M4 onset to E5 offset. Across the 12 intervention records, the median qualified orientation percentage was 14.52\% (interquartile range 9.19--26.62\%). NPC-available time ranged from 141.01 to 164.76~s per record. Control records were structurally unavailable for this target, rather than zero-valued observations under equivalent exposure. These descriptive values do not establish attention to spoken content or the effectiveness of companionship.

\paragraph{Pause records.}
The seven entries concerned P002, P003, P004, P009, and P011, all intervention-condition records. Four occurred within E1 and one each within E2, E3, and E4; none of the listed entries occurred within E5. Durations ranged from 2.50 to 5.625~s, totalling 27.50~s, and all seven entries marked resumption. The table contained no control pause entries. No listed pause overlapped an intervention-module interval or a pre-VAS 20-second window.

\section{User Study Questionnaires}
\label{app:study-questionnaires}

This appendix provides English translations of the Chinese questionnaire items administered in the study. Identifying administrative fields, such as participants' names, are omitted. The event-specific VAS-A item reproduces the in-VR prompt, while the remaining items reproduce the pre- and post-session questionnaire forms. These translations and module-specific adaptations should not be interpreted as independently validated versions of the measures.

\subsection{Background, Safety-Related Information, and MDAS}
\label{app:background-questionnaire-items}

Both groups completed the background and safety-related questions in Table~\ref{tab:background-safety-wording}. The response options below preserve the source form's categories rather than introducing additional history questions. All items were single-choice except the conditional prior-VR-symptoms item, which allowed multiple selections.

\begingroup
\small
\begin{longtable}{@{}>{\raggedright\arraybackslash}p{0.06\linewidth}>{\raggedright\arraybackslash}p{0.44\linewidth}>{\raggedright\arraybackslash}p{\dimexpr0.50\linewidth-4\tabcolsep\relax}@{}}
\caption{Background and safety-related questions and response options, translated from the pre-session forms.}\label{tab:background-safety-wording}\\
\toprule
Item & Question & Response options \\
\midrule
\endfirsthead
\multicolumn{3}{@{}l@{}}{\tablename~\thetable\ (continued)}\\
\toprule
Item & Question & Response options \\
\midrule
\endhead
\midrule
\multicolumn{3}{r@{}}{Continued on next page}\\
\endfoot
\bottomrule
\endlastfoot
1 & What is your age? & 18--24; 25--34; 35--44; 45--54; 55--64; 65 years or above. \\
\addlinespace
2 & What is your gender? & Man; woman. \\
\addlinespace
3 & What is your highest level of education? & High school or below; junior college; bachelor's degree; master's degree or above. \\
\addlinespace
4 & Have you previously received dental treatment? & Never; have received dental examinations and treatment. \\
\addlinespace
5 & Have you had an impacted wisdom tooth extracted? & Yes; no. \\
\addlinespace
6 & Approximately how long has it been since your most recent dental treatment? & Never received dental treatment; within one month; 1--3 months; 3--6 months; 6--12 months; one year or more. \\
\addlinespace
7 & Have you used VR (virtual reality) equipment before? & Never; once or twice; occasionally (1--3 times per month); frequently (at least once per week). \\
\addlinespace
8 & How often do you experience carsickness, seasickness, or other motion sickness? & Never; rarely; occasionally; often. \\
\addlinespace
9 & If you have used VR before, have you experienced any of the following discomforts? & No noticeable discomfort; dizziness; nausea; headache; eyestrain; loss of balance; other (please specify). Multiple selections allowed. \\
\addlinespace
10 & In the 24 hours before the experiment, have you taken medication that could affect heart rate, the nervous system, or physiological arousal? & No; yes. \\
\addlinespace
11 & In the six hours before the experiment, have you consumed caffeinated drinks or foods, such as coffee, strong tea, or energy drinks? & No; yes; unsure. \\
\addlinespace
12 & If you consumed caffeinated food or drinks, approximately how long ago was this? & None consumed; within one hour; 1--3 hours; 3--6 hours; six hours or more. \\
\addlinespace
13 & In the two hours before the experiment, have you undertaken vigorous exercise that noticeably increased your heart rate? & No; yes. \\
\addlinespace
14 & If you undertook vigorous exercise, approximately how long ago was this? & None undertaken; within 30 minutes; 30--60 minutes; 1--2 hours. \\
\end{longtable}
\endgroup

\paragraph{MDAS}
Both groups completed the five-item Chinese MDAS~\cite{humphris1995mdas}. The instruction asked participants to rate their anxiety in each situation below:
\begin{enumerate}[label=\arabic*., leftmargin=*]
  \item If you were going to have dental treatment tomorrow.
  \item When you are sitting in the waiting room awaiting treatment.
  \item If your teeth were going to be drilled.
  \item If your teeth were going to be polished.
  \item Having an anaesthetic injected into your gums (upper posterior tooth region).
\end{enumerate}
The form's response labels translate as 1 (not anxious), 2 (slightly anxious), 3 (mildly anxious), 4 (very anxious), and 5 (highly anxious). These translations retain the supplied labels, including the distinction between the second and third categories. Item scores were summed to give a total from 5 to 25.

\subsection{Current Anxiety, Event-Specific Anxiety, and VR Sickness}
\label{app:anxiety-sickness-questionnaire-items}

\paragraph{Baseline current VAS-A}
Before the tutorial, both groups answered: ``At this moment, how nervous or afraid do you feel?'' Responses were scored on a 0--100 range.

\paragraph{Event-specific VAS-A}
After each of E1--E5, both groups answered the same question: ``How nervous or afraid did the dental situation you just experienced make you feel?'' The response anchors were 0 (not nervous at all) and 100 (the strongest nervousness or fear imaginable). In the intervention condition, the associated module preceded the rating. For E1, this meant that VAS-A1 followed the M2 guidance period, which occurred after the logged E1 offset. Each item referred retrospectively to the event just experienced, not to anxiety predicted before the next event. The five ratings remained separate repeated observations rather than a summed scale.

\paragraph{Post-session current VAS-A}
After the VR run, both groups answered: ``At this moment, how nervous or afraid do you feel?'' The source form explicitly labels 0 as not nervous or afraid at all and 100 as the strongest nervousness or fear imaginable.

\paragraph{VRSQ}
Both groups were instructed to rate the severity of each discomfort symptom based on how they felt at that moment~\cite{kim2018vrsq}. The nine symptoms were general discomfort, fatigue, eyestrain, difficulty focusing, headache, fullness of the head, blurred vision, dizziness with eyes closed, and vertigo or a spinning sensation. Response labels were none, slight, moderate, and severe, coded 0, 1, 2, and 3, respectively.

The oculomotor score used the first four symptoms; the disorientation score used the remaining five. Their sums were divided by 12 and 15, respectively, and multiplied by 100. The total VRSQ score was the mean of the two subscale percentages. Questionnaire symptom scores remained distinct from observed adverse events.

\subsection{Intervention-Condition Questionnaire Items}
\label{app:intervention-questionnaire-items}

The intervention questionnaire contained five system diagnostic items and eight items repeated separately for each of M0--M4, yielding 40 module-specific items. No additional system-level AIM/IAM block or helpfulness, disruption, timing, and reuse matrix was administered.

\paragraph{System diagnostic items.}
The instruction asked participants to evaluate the statements based on the intervention support received throughout the VR experience. Responses ranged from 1 (completely disagree) to 5 (completely agree); intermediate options were labelled only 2, 3, and 4. These researcher-developed items concern intervention delivery and are reported separately, without a composite score. Their conceptual relationship to TFA does not establish that they form a validated TFA scale~\cite{sekhon2017acceptability}.

\begin{table}[htbp]
  \caption{English translations of the five system diagnostic items. The source form presents these as five rows of one response matrix.}
  \label{tab:system-diagnostic-wording}
  \small
  \begin{tabularx}{\linewidth}{@{}>{\raggedright\arraybackslash}p{0.07\linewidth}>{\raggedright\arraybackslash}X@{}}
    \toprule
    Item & Statement \\
    \midrule
    1 & The system provided interventions when I needed support. \\
    2 & The duration of each intervention was appropriate. \\
    3 & The interventions did not noticeably interrupt my experience of the dental situation. \\
    4 & During the experience, I could control pausing, continuing, or reducing the intensity of stimulation. \\
    5 & I understood why the system provided these interventions at the corresponding moments. \\
    \bottomrule
  \end{tabularx}
\end{table}

\paragraph{Module-specific AIM and IAM items.}
For each module, participants first evaluated their acceptance of the intervention and then its appropriateness. The module descriptions translated as M0, raising the left hand to pause; M1, breathing guidance and regulation; and M2, guidance and explanation of the surgical procedure. M3 was described as sensory regulation through visually blurring instruments and reducing sound volume. M4 was described as emotional support through having someone accompany the participant during the dental visit. The four AIM statements and four IAM statements were repeated unchanged across these five module referents~\cite{weiner2017implementation}. Table~\ref{tab:aim-iam-wording} translates the supplied Chinese statements rather than substituting the original English instrument wording.

\begin{table}[htbp]
  \caption{English translations of the four AIM and four IAM statements, each applied separately to M0--M4. The referent ``this intervention'' changes with the module.}
  \label{tab:aim-iam-wording}
  \small
  \begin{tabularx}{\linewidth}{@{}>{\raggedright\arraybackslash}p{0.10\linewidth}>{\raggedright\arraybackslash}X@{}}
    \toprule
    Item & Statement \\
    \midrule
    AIM 1 & I approve of this intervention. \\
    AIM 2 & I find this intervention appealing. \\
    AIM 3 & I like this intervention. \\
    AIM 4 & I welcome or am willing to accept this intervention. \\
    \addlinespace
    IAM 1 & This intervention fits the current situation. \\
    IAM 2 & This intervention is appropriate for the current need. \\
    IAM 3 & This intervention can be applied to the current problem. \\
    IAM 4 & This intervention is a good match for the target problem or population. \\
    \bottomrule
  \end{tabularx}
\end{table}

All module blocks used numeric response options from 1 to 5. The administered questionnaire labelled the endpoints completely disagree and completely agree, except in the M4 IAM block, which displayed numbers only.

For Figure~\ref{fig:system-module-evaluations}, each participant's four responses were averaged separately for each module and construct, without reverse scoring. M3 was rated as one combined visual-and-auditory intervention, not as separate M3.1 and M3.2 modules. The five diagnostic items remained separate.

\clearpage

\bibliographystyle{ACM-Reference-Format}
\bibliography{references}

\end{document}